\documentclass[preprint,12pt,authoryear]{elsarticle}
\usepackage[T1]{fontenc}
\usepackage{lmodern}
\usepackage{amsmath,amssymb,amsthm}
\usepackage{booktabs}
\usepackage{adjustbox}
\usepackage{graphicx}
\usepackage[dvipsnames]{xcolor}
\usepackage{algorithm}
\usepackage{algpseudocode}
\usepackage[colorlinks=true,linkcolor=blue!50!black,citecolor=blue!50!black,urlcolor=blue!50!black]{hyperref}
\usepackage[capitalise,noabbrev]{cleveref}
\makeatletter\def\theHALG@line{\thealgorithm.\arabic{ALG@line}}\makeatother
\crefformat{appendix}{#2#1#3}\Crefformat{appendix}{#2#1#3}

\newtheorem{definition}{Definition}
\newtheorem{proposition}{Proposition}
\newtheorem{remark}{Remark}

\newcommand{\Q}{\mathbb{Q}}
\newcommand{\Z}{\mathbb{Z}}
\newcommand{\PGL}{\mathrm{PGL}}
\newcommand{\GL}{\mathrm{GL}}
\newcommand{\SL}{\mathrm{SL}}
\newcommand{\M}{\mathcal{M}}
\newcommand{\Res}{\operatorname{Res}}
\newcommand{\hmin}{h_{\min}}
\newcommand{\halg}{h_{\mathrm{alg}}}

\makeatletter
\def\ps@pprintTitle{\let\@oddhead\@empty\let\@evenhead\@empty\let\@oddfoot\@empty\let\@evenfoot\@empty}
\makeatother

\begin{document}

\begin{frontmatter}

\title{Where models of minimal height are hard to find:\\ cubic rational maps with a multiple fixed point}

\author{Elira Shaska}
\ead{elirashaska@oakland.edu}
\affiliation{organization={Department of Computer Science, Oakland University},
            city={Rochester},
            state={MI},
            postcode={48309},
            country={USA}}

\begin{abstract}
Given a rational point of the moduli space of cubic rational maps, we seek a map in its class with the smallest integer coefficients. A recent
algorithm \citep{m3} combines an exact descent and minimisation of the resultant with a reduction and a local search; it is nearly exact where the
minimal height is known. Beyond that range, a certified deeper search shows that the instances in which a smaller model lies beyond its reach are
not random: a classifier predicts them from the six invariants of the point, and asking what it relies on leads to an exact algebraic surface, the
maps with a triple fixed point. Invariant theory and the stability hypothesis of reduction theory \citep{hutz-stoll} then identify the larger locus that
contains it: the hypersurface $\xi_1^3=6\xi_2^2$ of maps with a multiple fixed point, on which the fixed-point quartic is not stable. It holds most
of the hard instances. On it the two invariants of the fixed-point quartic are $(\xi_1,\xi_2)=(6m^2,6m^3)$ for an integer $m\ge0$; we
parametrise the integer points of the surface, and show that on its generic part the sign of $\xi_5$ is the sign of the cubic coefficient at the
parabolic fixed point. A covariant defined on the locus, added to the reduction,
resolves only a minority of the hard instances: they appear to be mainly a matter of the search's reach. These findings lead to a certified structured search
that never returns a larger model than the original algorithm and finds most of the smaller models within reach, also at heights not used to design
it, at a small multiple of its running time.
\end{abstract}


\end{frontmatter}

\section{Introduction}
\label{sec:intro}

A rational map $\phi\colon\mathbb P^1\to\mathbb P^1$ of degree $3$ over $\Q$ is determined, up to change of coordinates, by six integer
invariants $\xi_0,\dots,\xi_5$, which place its conjugacy class at a point of a moduli space $\M_3$ \citep{jsc}. The forward direction, from a map to
its invariants, is a polynomial evaluation. The inverse direction asks, given a point $P\in\M_3(\Q)$, for a map with rational coefficients in the
class $P$ whose coefficients are as small as possible; the smallest attainable maximum $|c_i|$ is the minimal naive height $\hmin(P)$. It is
computable in principle, by enumerating the primitive coefficient vectors up to the height of any known model and testing each for membership in
the class, but this is impractical beyond small heights, and we are not aware of an efficient general method.

The minimal naive height is not the only notion of a smallest model. One can minimise the resultant \citep{bruin-molnar}, or the height among the models of minimal resultant \citep{hutz-stoll}; both have
arithmetic meaning. We study $\hmin$ because it is the size of the smallest rational model of the
class, the one that is listed and computed with: the database \citep{jsc} enumerates maps by naive height, the cost of iterating a map or computing
with it grows with the size of its coefficients, and $\hmin(P)$ can be strictly smaller than the height of every model of minimal resultant
\citep[Ex.~7.2]{m3}, in which case the reduced models of reduction theory \citep{hutz-stoll} do not give it. At small heights this is rare: among the generic classes
with $\hmin\le4$ it happens for $0.15\%$, about one class in $700$, almost always one step at the prime $2$ from the minimal-resultant locus.

A natural method has two parts. Exact steps construct a model and make its resultant minimal: an explicit descent \citep[Prop.~3.7]{m3} gives a
map over $\Q$ from a fine moduli point, and its resultant is then minimised prime by prime \citep{bruin-molnar}. A search over changes of
coordinates then makes the model small: a reduction at infinity in the spirit of reduction theory \citep{hutz-stoll}, followed by a local search. The algorithm
\citep[Alg.~7.1]{m3}, which we call Algorithm~A and specify completely in \cref{app:algA}, combines them with a local search of depth three. Within the range
where $\hmin$ is known it is nearly exact: the original implementation returns a model of minimal height for all but one of its $7\,409$ test
classes with $\hmin\le4$ \citep[\S7]{m3}, and ours does so for all $461$ points of our census samples whose minimal height is known
(\cref{sec:data}), though not for every class, as the example $P_0$ of \cref{sec:problem} shows. That range is small: only $0.04\%$ of the fine points
of the census have $\hmin\le4$ \citep[\S7]{m3}, and for the others the height of the best model found is an upper bound for $\hmin$,
nothing more.

This paper studies the problem beyond that range, on rational points whose minimal height is unknown. The setting suits a computational study.
Every candidate model is checked exactly and cheaply by recomputing its invariants, so any smaller model is a certificate; exact values of $\hmin$
are known for $24.4$ million classes from an exhaustive enumeration of maps with small coefficients; and the census \citep{m3} supplies the moduli
points on which the problem is open. As a reference we use a certified search of depth five, and we call a point a hard instance when that search
finds a smaller model than Algorithm~A.

Our first finding is that the hard instances are not random. They are rare, about one point in seventy, but they are predictable from the six
invariants alone, and asking what a classifier relies on leads to an exactly defined locus: the surface $\xi_1=\xi_2=0$ of maps with a fixed point
of multiplicity three, which holds most of the hard instances at small heights. The surface is a cuspidal cylinder with an explicit integer
parametrisation, and its two halves, distinguished by the sign of $\xi_5$, behave very
differently. We show (\cref{prop:sign}), by a computation in a normal form, that on its generic part this sign is the sign of the cubic coefficient at the parabolic fixed point, so
that the two halves are the maps whose triple fixed point attracts along real and along imaginary directions.

Invariant theory and the stability hypothesis of reduction then lead from the surface to the hypersurface $\xi_1^3=6\xi_2^2$ of maps with a
multiple fixed point, of which the surface is the most degenerate part. It is exactly where the fixed-point quartic stops being stable in the sense of reduction theory
\citep{hutz-stoll}: the quartic is strictly semistable at a double fixed point and unstable at a triple one. On the hypersurface the two invariants of
the fixed-point quartic are $(\xi_1,\xi_2)=(6m^2,6m^3)$ for an integer $m\ge0$, and it holds the large majority of the hard instances, those with a triple fixed
point at small heights and increasingly those with a double one as the height grows. Since this is also where the reduction step of Algorithm~A
loses one of its covariants, we tested whether the hard instances are an artefact of the reduction: adding a covariant that is defined there, the
covariant point of the form of period two, resolves only a minority of them. The smaller models lie just beyond the reach of the local search, and
on the surface the paths to them share a recurring shape, a few conjugations at $2$ and $3$ followed by the elliptic element $ST$.

These findings lead to a certified structured search, Algorithm~A$^+$: composite moves read off the improving paths, and a deeper search on the
points that a rule selects, either a classifier or the exact condition that a fixed point is multiple. Every candidate is verified, so the result is
never larger than Algorithm~A's. The search finds most of the smaller models within reach of the reference, also on two ranges of heights that were
used neither to train nor to design it, at a small multiple of the running time of Algorithm~A.

Following an approach used elsewhere in mathematics \citep{davies-2021}, the paper is also a case study of machine learning as a diagnostic tool for an exact algorithm: a fitted model pointed to the surface where the hard
instances concentrate, invariant theory placed it within a larger locus, and this led to a better search, in which a learned component is safe because every output is
certified. All data are generated by our own programs \citep{minmodels}, which implement the invariants, automorphism strata, canonical representatives, census and
Algorithm~A of the earlier work \citep{jsc,m3}.

\Cref{sec:problem} recalls the problem and summarises Algorithm~A, which \cref{app:algA} specifies in full, and \cref{sec:related} discusses related
work. \Cref{sec:setup} describes the data, the reference search and the features. \Cref{sec:misses} measures and predicts the hard instances,
\cref{sec:surface} locates them on the locus of multiple fixed points, and \cref{sec:why} indicates that they are mainly a matter of the reach of the
search. \Cref{sec:method} presents Algorithm~A$^+$, \cref{sec:eval} evaluates it, and \cref{sec:conclusion} discusses the limitations and concludes.

\section{Background and problem formulation}
\label{sec:problem}

A cubic map over $\Q$ is a primitive vector $c=(c_0,\dots,c_7)\in\Z^8$, the coefficients of $f_0=c_0x^3+\dots+c_3y^3$ and
$f_1=c_4x^3+\dots+c_7y^3$, with $\Res(f_0,f_1)\neq0$ and first nonzero entry positive; its naive height is $h(c)=\max_i|c_i|$.
$\GL_2$ acts by conjugation, and two maps are conjugate over $\overline{\Q}$ if and only if their invariants $\xi_0,\dots,\xi_5$, of
weights $(2,2,3,3,4,6)$, agree up to weighted scaling \citep{jsc}. Conjugacy over $\Q$ is finer in general: a moduli point with nontrivial
automorphism group can carry twists, rational maps with the same invariants that are not conjugate over $\Q$. For a fine point with trivial
automorphism group, as are all points in our samples, all rational maps representing the point are conjugate over $\Q$ \citep[Lemma~8.3]{m3}, so the
two notions agree. We represent a class by its canonical representative \citep[\S2.2]{m3}, and its moduli height is
\[
H(P)=\max_i|\xi_i|^{1/w_i}.
\] A point is \emph{fine} if its class contains a map over $\Q$ \citep[Thm.~3.3]{m3}.

\begin{definition}
For a fine $P\in\M_3(\Q)$,
\[
\hmin(P)=\min\{h(c): c \text{ a map over }\Q\text{ in the class }P\}.
\]
\end{definition}

Algorithm~A \citep[Alg.~7.1]{m3} is specified in full in \cref{app:algA}. In brief: (1) an explicit descent gives a map
with invariants $P$; (2) the resultant is minimised prime by prime in the Bruhat--Tits trees, which is exact by the convexity of
$\operatorname{ord}_p\Res$ on the tree \citep{rumely,bruin-molnar}; (3) the models of equal minimal resultant nearby in the trees are collected;
(4) each is reduced at infinity, by moving covariant points into the fundamental domain of $\SL_2(\Z)$ and searching short words \citep{hutz-stoll};
(5) a local search of depth $3$ over eleven moves, $z\mapsto z\pm1$, $-1/z$, $-z$ and seven conjugations of determinant $2$ and $3$, keeps the best
model found and repeats while the height decreases; (6) the smallest model found is returned. Steps 1--2 are exact; steps 3--5 are search, and the
height $\halg$ of the answer is an upper bound for $\hmin$. \Cref{app:algA} also states how our implementation reads each heuristic step. \Cref{sec:checks} compares it with a first-improvement reading of step 5.

A smaller model can be out of reach for three documented reasons. The model of minimal naive height may not have minimal resultant
(\citealp[Ex.~7.2]{m3}); it may have minimal resultant but lie beyond the radius of the search (the class $[-48:2:0:48:-48:-256]$, the
one test class \citep[\S7]{m3} on which the original implementation returns height $3$ where $\hmin=2$; ours reaches height $2$ in step 4); or it may be reachable only through a prime the search does not visit
(\citealp[\S7]{m3}, and the class $P_0=[2520:6:6:29232:-97776:16003008]$, of minimal height $4$, known from exhaustive enumeration \citep{jsc}, for which
Algorithm~A returns height $5$; the conjugation between the two models has determinant $-14$, and the path between them is $z\mapsto-z$, a $7$-adic
step, $z\mapsto-1/z$, a $2$-adic step, which a search moving only at $2$ and $3$ cannot follow at any depth).

\section{Related work}
\label{sec:related}

The background of Algorithm~A consists of models of minimal resultant \citep{bruin-molnar}, the convexity of the resultant on the Bruhat--Tits
tree and the minimal-resultant locus \citep{rumely}, and reduction at infinity \citep{hutz-stoll}, which builds on the reduction theory of binary
forms \citep{cremona-stoll}; the general setting is that of arithmetic dynamics \citep{silverman-ads}. Reduction theory finds a model of smallest height among the models of minimal resultant \citep[Def.~2.1]{hutz-stoll}; $\hmin$ is the minimum over all rational
models, which can lie off the minimal-resultant locus \citep[Ex.~7.2]{m3}, so this does not solve our problem; it also requires the binary
form attached to the map to be stable, with no root of multiplicity at least half its degree, using the forms of periodic points of higher period
when the fixed-point form is not \citep[\S\S3, 6]{hutz-stoll}. For binary forms, neither Julia reduction nor hyperbolic reduction reaches the minimal height in general: a small further shift and a rescaling are usually needed, and a neurosymbolic model whose symbolic layers search for them has been proposed \citep{kotsireas-shaska}. The
moduli space of cubic maps and its invariants \citep{west,jsc}, and the fine/coarse question, the census and Algorithm~A \citep{m3}, are the starting
point of this paper. The local picture at a parabolic fixed point used in \cref{prop:sign} is the Leau--Fatou flower theorem \citep{milnor-book}.

Choosing, per instance, which algorithm or how much computation to spend is the algorithm selection problem \citep{rice-1976}, studied for combinatorial search \citep{kotthoff-2014} and made practical by portfolios such as
SATzilla \citep{satzilla}; learning has also been used inside exact solvers \citep[surveyed in][]{bengio-lodi-2021}, for branching in mixed-integer programming
\citep{gasse-2019} and for satisfiability \citep{neurosat}. In computer algebra, learned heuristics choose the variable ordering for
cylindrical algebraic decomposition \citep{huang-2014} and the selection strategy in Buchberger's algorithm \citep{peifer-2020}. Algorithm~A$^+$ is an
instance of this pattern with two differences: every accepted output is certified by exact computation, so errors of the classifier cost time or a missed improvement but never a worse
answer, and the learning was also used to locate where the instances are hard.

Fitted models have guided mathematicians to structure, by asking which inputs a model relies on, in
knot theory and representation theory \citep{davies-2021}; \cref{sec:surface} follows the same route to an exact locus. Generate-and-verify with an
exact verifier is the pattern of recent work \citep{funsearch,alphageometry}; learning from integers and their digits has also been studied \citep{charton}. On moduli
data, classifiers have learned automorphism strata of cubic maps \citep{jsc} and loci of genus-two curves with split Jacobian \citep{ss-jac}. In
arithmetic dynamics, a genetic algorithm has generated extreme examples of rational maps \citep{hutz-ga}; for some of them, a model of smallest height
among the models of minimal resultant \citep{hutz-stoll} was not reasonably computable.

\section{Experimental setup}
\label{sec:setup}

\subsection{Data}
\label{sec:data}

As for the database \citep[\S6.1]{jsc}, we enumerate all $104\,371\,000$ primitive maps with $|c_i|\le5$, independently of the programs
used there, and group them by canonical moduli point, which gives $\hmin$ exactly for the $24\,436\,749$ classes with $\hmin\le5$; the classes with
$\hmin\le4$ are those of the database. For every generic class with $\hmin\le4$ we also record whether some model of minimal height has
minimal resultant. In this paper the exact labels serve for the comparison with the minimal-resultant locus in the introduction, and to check Algorithm~A
where the minimal height is known: $461$ points of the three census samples below have a known minimal height ($451$ with $\hmin\le4$ and $10$
with $\hmin=5$), and Algorithm~A returns it on every one of them.

The census \citep[\S5]{m3} is the list of the rational points of $\M_3$ by moduli height. We generated it
ourselves, with our programs \citep{minmodels}, which implement the canonical representatives and the fineness criterion \citep{m3}; to $H\le5$ it has $12\,438\,121$ points, of which $9\,215\,604$ are fine \citep[as in][Table~2]{m3}. Algorithm~A applies to fine generic points with $\xi_0m_{12}\neq0$, where
\[
m_{12}=\tfrac13\xi_0^2\xi_1+\xi_0\xi_4-\xi_3^2;
\]
points with $\xi_0m_{12}=0$ are fine automatically \citep[Thm.~3.3(2)]{m3} but need a
different construction, and are excluded ($15.1\%$ of the fine generic points with $H\le5$, almost all with $\xi_0=0$). We ran
Algorithm~A on all $1\,725$ eligible points with $H\le2$ and on $20\,000$ drawn at random in each of the shells $2<H\le3$, $3<H\le4$,
$4<H\le5$: $61\,725$ points, all processed without failure. Every model returned by Algorithm~A was verified to lie in its class
by recomputing its canonical invariants: all $61\,725$ here, and all $10\,000$ in each of the samples at $5<H\le6$ and
$6<H\le7$ used later. The median of $\halg$ by shell, over all points of the sample, is $18$,
$30$, $50$, $78$, against $18$, $31$, $54$, $72.5$ on a smaller sample \citep[Table~9]{m3}; the two are computed on different samples and with
independent implementations of the heuristic steps 3--5, and we have not traced the difference further.

Two further samples test generalisation (\cref{sec:general}). At $5<H\le6$ we drew $10\,000$ points uniformly, by
reservoir sampling, from the $29\,779\,198$ fine generic points with $\xi_0m_{12}\neq0$ of the census; at $6<H\le7$, $10\,000$ such points from a uniform
$0.1\%$ thinning of our enumeration of the census to $H\le7$. Neither was used to train the classifiers or to design the targeted repair, and the
method of \cref{sec:method} was fixed before either was evaluated.

\subsection{Reference search and hard instances}
\label{sec:reference}
From the model after step 4 of Algorithm~A, where its own local search begins, we search breadth-first to depth $5$ over
$z\mapsto z\pm1$, $-1/z$, $-z$ and the conjugations by $\left(\begin{smallmatrix}p&a\\0&1\end{smallmatrix}\right)$ and
$\left(\begin{smallmatrix}1&0\\0&p\end{smallmatrix}\right)$ at $p=2,3$ and at every prime $p\le13$ with $p^2\mid\Res$, with duplicate detection,
a budget of $2\cdot10^6$ nodes and a timeout of $600$\,s per point. The best model is verified by recomputing its canonical invariants whenever it is smaller than the model after step 4, as it is at every miss, and the best height at each depth and the path of moves to the best model are recorded; every miss below is certified by a
verified model (\cref{sec:rates}). We ran this on all $61\,725$ census
points of \cref{sec:data}, and call it the \emph{reference search}. Its results are upper bounds for $\hmin$, not certificates of minimality.

For a point, let $b_3$ and $h_{\mathrm{best}}$ be the smallest heights found by the reference search within depth $3$ and within depth $5$ of the
step-4 model. The point is a \emph{miss}, or a \emph{hard instance}, when a deeper search finds a smaller model than the local search of
Algorithm~A; it is a \emph{greedy miss} when that happens already within depth $3$, and a \emph{genuine miss} when the smaller model lies deeper:
\[
\begin{aligned}
&\text{miss:} && h_{\mathrm{best}}<\halg,\\
&\text{greedy miss:} && b_3<\halg,\\
&\text{genuine miss:} && h_{\mathrm{best}}<\min(\halg,b_3).
\end{aligned}
\]
A genuine miss is a point for which the reference search finds an improvement beyond those reachable in a
single depth-$3$ neighbourhood of the model after step 4. Since step 5 repeats its depth-$3$ search, Algorithm~A itself can reach
beyond a single neighbourhood; the points where it does are not misses. A point can be both a greedy and a
genuine miss, when there is an improvement within depth $3$ and a better one deeper. Since step 5 returns the best model within depth $3$ with its
own moves, a greedy miss requires a move step 5 does not have: the reference search also moves at the primes $5$ to $13$ that divide the resultant to
the second power. Recomputed with exactly step 5's eleven moves, the genuine-miss label changes on $54$ of the $61\,725$ points. We keep the
reference's definition for the analyses of \cref{sec:misses,sec:surface}, which concern where the reference search finds improvements, and measure
Algorithm~A$^+$ directly against all misses. All rates of misses are relative to the reference search, and therefore lower bounds for the rate at which $\halg>\hmin$. The two searches are
not nested: step 5 restarts its depth-$3$ search from every improvement, so its path can be longer than five moves, and Algorithm~A's answer is
smaller than the reference's best on $12$ of the $61\,725$ points at $H\le5$ and on $1$ and $3$ of the $10\,000$ points of the held-out samples.
This does not affect the miss counts, which are one-sided, and all recovery rates below are measured on points where the reference search beats
Algorithm~A.

\subsection{Two checks on Algorithm~A and the reference}
\label{sec:checks}
A first-improvement reading of step 5, which searches one depth at a time and moves, at the first depth containing a smaller model, to the last
such model it encounters, not necessarily the smallest, misses on $1\,646$ of the
$61\,725$ points ($2.7\%$), and on $336$ and $421$ of the $10\,000$ points of the held-out samples of \cref{sec:general}, against $130$ and $188$ for
Algorithm~A. Below height $5$, $741$ points are misses of both, Algorithm~A misses on $95$ points where the variant succeeds, and of the two answers
Algorithm~A's is the smaller on $947$ points and the variant's on $107$. Both rules cost the same, about $0.015$\,s per point. The comparison shows
that the choice made within the radius of the search matters as much as the radius, and it is the reason we describe Algorithm~A's step 5 exactly as published.

The reference search is largely insensitive to the starting model. Any two rational models of a class with trivial automorphism group differ by a unique
element of $\PGL_2(\Q)$ \citep[Lemma~8.3]{m3}, so taking another rational point on the descent conic is the same as starting from a rational conjugate of the descended model. For
$2\,000$ points ($500$ per shell) we ran steps 1--4 and the deep search from the descended model and from two random rational conjugates of it.
The height after step 4 is the same from all three starts on $99.7\%$ of the points, and the best model found by the deep search on $99.9\%$:
steps 2--4 largely erase the starting model, and the reference is stable.

\subsection{Features and models}
\label{sec:features}
The invariant-only models (sets A--D below) take as input the canonical moduli point only; the predictor used after
Algorithm~A has run, and in Algorithm~A$^+$, adds $\log\halg$. The features of the moduli point are: signed logarithms and weight-zero coordinates
of the $\xi_i$; $\log H$; $m_{12}$, a principal minor of the Gram matrix of three covariant quadratics that cuts out the involution locus
$L_2=\{\xi_5=m_{12}=0\}$ and enters the descent obstruction \citep[Lem.~3.1, Thm.~3.3]{m3}, and the resultant $I_6(P)$ of the canonical point
\citep[Eq.~(3)]{m3}, as signed logarithms; the valuations at $2,3,5,7$ of the $\xi_i$, of $3m_{12}$ and of $768\,I_6$ (both integers);
residues modulo $8$ and $9$; and two signs. The sets A (geometry: sizes and weight-zero coordinates), B (A with $m_{12},I_6$),
C (B with the valuations) and D (all) form a ladder.

\section{Hard instances}
\label{sec:misses}

\subsection{How often a smaller model lies beyond the local search}
\label{sec:rates}
\begin{table}[ht]
\centering
\caption{Hard instances by shell (definitions in \cref{sec:reference}). Improvement factor: median of $\halg/h_{\mathrm{best}}$ over the
misses.}
\label{tab:misses}
\small\setlength{\tabcolsep}{4pt}
\begin{adjustbox}{max width=\textwidth}
\begin{tabular}{lccccc}
\toprule
shell & $H\le2$ & $2<H\le3$ & $3<H\le4$ & $4<H\le5$ & all\\
\midrule
points & 1\,725 & 20\,000 & 20\,000 & 20\,000 & 61\,725\\
misses & 25 (1.4\%) & 265 (1.3\%) & 303 (1.5\%) & 243 (1.2\%) & 836 (1.4\%)\\
\quad greedy (within depth 3) & 0 & 16 & 24 & 23 & 63\\
\quad genuine (need depth 4--5) & 25 (1.4\%) & 253 (1.3\%) & 281 (1.4\%) & 223 (1.1\%) & 782 (1.3\%)\\
median improvement factor & 1.11 & 1.13 & 1.11 & 1.09 & \\
search completed within budget & 98.4\% & 96.1\% & 96.1\% & 95.1\% & 95.8\%\\
\bottomrule
\end{tabular}
\end{adjustbox}
\end{table}

For $836$ of the $61\,725$ points ($1.35\%$; \cref{tab:misses}) the reference search finds a smaller model than Algorithm~A. Every miss is certified by the reference search's verified best model. Almost all the missing improvements lie beyond the radius of step 5:
$782$ are found at depth $4$ or $5$ ($424$ and $358$), and only $54$ within depth $3$, all through moves step 5 does not have; $101$ of the
improving paths use a prime above $3$, at which step 5 never moves. The miss rate is $1.2$--$1.5\%$ in every shell, but this flatness is a
coincidence of opposite trends, analysed in \cref{sec:surface}: most genuine misses at small heights lie on a surface that thins out with the
height, while the rate away from it grows. The sample is stratified by shell, so pooled figures are averages over the sample, not over the census,
whose eligible points with $H\le5$ lie to $88\%$ in the outer shell. Weighted by the census, the miss rate is $1.25\%$ and the genuine-miss rate
$1.15\%$ (against $1.35\%$ and $1.27\%$ in the sample), and the surface of \cref{sec:surface} holds $48\%$ of the genuine misses rather than $71\%$.
Per-shell figures, which do not depend on the weighting, are reported throughout, and the classifiers are trained on the stratified sample. The
improvements are modest, typically $10$--$15\%$. None of the genuine-miss labels comes from a search that completed fewer than three depths, and
$16$ of the $782$ come from searches that reached the node budget; since a label requires a smaller model, truncation can only hide misses, not
create them. 

\subsection{The hard instances are predictable}
\label{sec:predict}
Gradient-boosted trees (\texttt{HistGradientBoosting}, balanced class weights), 5-fold
stratified cross-validation on the $61\,725$ points of \cref{sec:misses}, with two labels: all misses, and genuine misses only. Intervals are
bootstrap intervals over the points with the out-of-fold scores held fixed, which understates the variance due to refitting.

\begin{table}[ht]
\centering
\caption{Predicting the hard instances: AUC of out-of-fold predictions with $95\%$ bootstrap intervals, and average precision (AP);
base rates $1.35\%$ and $1.27\%$. Rows A--D use the moduli point only; the last row uses only the primes at which the reference search moves.}
\label{tab:beatable}
\begin{adjustbox}{max width=\textwidth}
\begin{tabular}{lcc}
\toprule
features & all misses ($836$) & genuine misses ($782$)\\
\midrule
$\log H$ only & 0.74 [0.72, 0.76] & 0.74 [0.72, 0.76], AP 0.10\\
A: geometry & 0.93 [0.92, 0.94] & 0.93 [0.92, 0.94], AP 0.36\\
B: A $+\,m_{12},I_6$ & 0.93 [0.92, 0.94] & 0.93 [0.91, 0.94], AP 0.35\\
C: B $+$ valuations & 0.94 [0.93, 0.95] & 0.93 [0.92, 0.95], AP 0.49\\
D: all & 0.94 [0.93, 0.95], AP 0.48 & 0.93 [0.92, 0.94], AP 0.48\\
\midrule
$\log\halg$ (output of the algorithm) & 0.67 [0.66, 0.68], AP 0.021 & 0.67 [0.66, 0.68], AP 0.020\\
D $+\log\halg$ & 0.95 [0.95, 0.96], AP 0.51 & 0.95 [0.94, 0.96], AP 0.52\\
primes at which the search moves & 0.51, AP 0.01 & 0.55, AP 0.01\\
\bottomrule
\end{tabular}
\end{adjustbox}
\end{table}

The moduli point predicts the hard instances with AUC $0.94$ and average precision $0.48$, a $36$-fold enrichment over the base rate of
$1.35\%$, against AUC $0.67$ from the size of the answer (\cref{tab:beatable}): the invariants carry information about the hard instances beyond the
fact that larger answers leave more room. The feature sets A to D do not differ beyond their intervals, which understate the variance: the geometry
features alone appear to suffice. The genuine misses, whose improvements lie beyond a single depth-$3$ neighbourhood of the model after step 4, are predicted
with AUC $0.93$, and $0.95$ in the realistic setting where the decision is taken after Algorithm~A has run and $\halg$ is known.

The label is defined by the deep search, which moves at more primes when more small primes
divide the resultant; since the features contain valuations at $2,3,5,7$, the classifier could in principle be predicting the width of the
search rather than anything about the class. The primes at which the search moves predict the misses at chance level (AUC $0.51$--$0.55$, AP
equal to the base rate; the number of models examined is a function of that move set and adds nothing). A sharper test holds the width fixed:
at the $46\,986$ points where the search moves only at $2$ and $3$, every search uses the same eleven moves, the same depth and the same budget,
and none was truncated. These points contain $671$ of the $782$ genuine misses, and on them the classifier on the moduli point and $\log\halg$
predicts the genuine misses with AUC $0.961$ and AP $0.56$, as well as on all points. (The genuine-miss rate is lower among the other points,
$0.75\%$ against $1.43\%$, but $17.5\%$ of their searches reached the node budget, so that rate is a lower bound and we draw no conclusion
from the difference.)

Grouped permutation importance of the genuine-miss classifier on the $\{2,3\}$ subset, over five
stratified folds with disjoint feature groups (held-out AUC $0.95$--$0.97$ in every fold), points to the position of the moduli point in the weighted
projective space: permuting the six weight-zero coordinates $\xi_i/H^{w_i}$ lowers the AUC by $0.12$ ($0.08$--$0.17$ across folds), permuting
$\log\halg$ by $0.06$ ($0.05$--$0.09$), permuting the valuations at $2$ by $0.01$ ($0.005$--$0.013$), and permuting the signed sizes of the $\xi_i$, the valuations at $3,5,7$, the sizes
of $m_{12}$ and $I_6$, the residues or the signs by at most $0.005$, within the noise across folds. Among single coordinates, $\xi_2/H^3$ matters most, followed by
$\xi_5/H^6$ and $\xi_1/H^2$, and the misses lie at the small end of $\xi_1/H^2$ and $\xi_2/H^3$.

\section{Where the hard instances lie: maps with a multiple fixed point}
\label{sec:surface}

The hard instances concentrate on the maps with a multiple fixed point, the hypersurface $\xi_1^3=6\xi_2^2$ on which the fixed-point quartic is
not stable. \Cref{sec:double} describes this locus and how many of the hard instances it holds, \cref{sec:triple,sec:halves} its most degenerate
part, the surface of triple fixed points, where they are densest at small heights, and \cref{sec:found} how the locus was found;
\cref{sec:surfheight} follows the surface and the rest of the locus across the heights.

\subsection{The locus of multiple fixed points}
\label{sec:double}
The roots of the fixed-point quartic $I$ are the fixed points of $\phi$, and a fixed point is multiple exactly when its multiplier is $1$, since
the derivative of $\phi(z)-z$ vanishes there. With the normalisation of the transvectants used here \citep{m3}, $\xi_1=I/6$ and $\xi_2=J/72$ for the
classical invariants $I$ and $J$ of the fixed-point quartic \citep[in the notation of][\S9]{janson}, whose discriminant is $(4I^3-J^2)/27$
\citep[Eq.~(9.7)]{janson}; so the fixed-point quartic has a repeated root exactly when
\[
\xi_1^3=6\xi_2^2,
\]
and a root of multiplicity at least three exactly when $\xi_1=\xi_2=0$, since a binary form of degree $n$ has all its invariants zero if and only if it has a root of multiplicity greater than
$n/2$ \citetext{\citealp[Thm.~5.4]{janson}; \citealp[Cor.~1]{curri}}. The hypersurface $\xi_1^3=6\xi_2^2$ therefore consists of the maps with a fixed point of
multiplier $1$: of multiplicity two, except on its most degenerate part, the surface $\xi_1=\xi_2=0$ of \cref{sec:triple}, where three or four
fixed points have merged. A fixed point of multiplicity $m$ has $m-1$ attracting directions \citep[\S10]{milnor-book}, one at a double fixed point
and two at a triple one. It is
also exactly the locus where the fixed-point quartic is not stable in the sense of reduction theory \citep[\S3]{hutz-stoll}, where a binary form of degree $n$ is called stable when no
root has multiplicity at least $n/2$: for a quartic, when its roots are distinct. At a double fixed point the quartic is
still semistable, every root having multiplicity at most $n/2$ \citep[Thm.~2]{curri}, and its invariants do not all vanish; on the surface, the null
cone, it is unstable \citep[Thm.~5.4]{janson}. All quartics with a double root and no triple one have the same moduli point, the unique strictly
semistable point $[I:J]=[1:-2]$ of the moduli space of binary quartics \citep[Cor.~1, Lemma~5 and Table~1]{curri}; in our normalisation this is
exactly the relation $\xi_1^3=6\xi_2^2$, an independent check of the constant. As a check of the conventions, the
multiplicity class read off the invariants agrees with the exact factorisation of $I$ on $4\,572$ maps, $1\,031$ of them with a double fixed point and
$671$ with a triple or higher one.

On the hypersurface the two invariants of the fixed-point quartic take explicit values. Comparing valuations in $\xi_1^3=6\xi_2^2$ gives $3v_p(\xi_1)=2v_p(\xi_2)+[p\in\{2,3\}]$ for
every prime $p$, so the solutions are $(\xi_1,\xi_2)=(6m^2,\pm6m^3)$ with $m\in\Z$. In the canonical representative $\xi_2\ge0$, since $\xi_2$ is
its first coordinate of odd weight, so that on the hypersurface
\[
(\xi_1,\xi_2)=(6m^2,\,6m^3),\qquad m\ge0,
\]
with $m=0$ on the surface. As $|\xi_1|\le H^2$, a double fixed point with parameter $m\ge1$ needs $H\ge m\sqrt6$: no point with $H\le2$ has
one, and in all our samples, which stop at $H=7<3\sqrt6$, $m$ is $1$ or $2$, that is, $(\xi_1,\xi_2)=(6,6)$ from $H\approx2.45$ on and $(24,48)$ from
$H\approx4.90$ on. The example $P_0$ of \cref{sec:problem} is such a point, with $m=1$.

\begin{table}[ht]
\centering
\caption{Genuine misses by the multiplicity of the fixed points: shares of the points, genuine-miss rates, and the share of the genuine misses
that lie on the locus of multiple fixed points. Weighted by the census, the locus holds $13.9\%$ of the points with $H\le5$ and $84\%$ of their
genuine misses.}
\label{tab:multiplicity}
\small\setlength{\tabcolsep}{4pt}
\begin{adjustbox}{max width=\textwidth}
\begin{tabular}{lccccccc}
\toprule
 & \multicolumn{2}{c}{share of points} & \multicolumn{3}{c}{genuine-miss rate} & \\
\cmidrule(lr){2-3}\cmidrule(lr){4-6}
sample & double & triple & simple & double & triple & on the locus\\
\midrule
$H\le2$ & 0\% & 18.2\% & 0.07\% & --- & 7.6\% & 24/25\\
$2<H\le3$ & 15.3\% & 7.3\% & 0.10\% & 0.36\% & 15.6\% & 237/253\\
$3<H\le4$ & 10.1\% & 4.8\% & 0.21\% & 2.0\% & 21.3\% & 246/281\\
$4<H\le5$ & 11.0\% & 2.7\% & 0.22\% & 4.0\% & 17.9\% & 185/223\\
$H\le5$, all & 11.8\% & 5.3\% & 0.18\% & 1.9\% & 16.9\% & 692/782 (88\%)\\
\midrule
$5<H\le6$ (held out) & 7.6\% & 1.8\% & 0.32\% & 8.1\% & 14.8\% & 88/117 (75\%)\\
$6<H\le7$ (held out) & 6.9\% & 1.5\% & 0.37\% & 14.4\% & 21.9\% & 132/166 (80\%)\\
\bottomrule
\end{tabular}
\end{adjustbox}
\end{table}

\begin{figure}[ht]
\centering
\IfFileExists{fig_multiplicity.pdf}{\includegraphics[width=\textwidth]{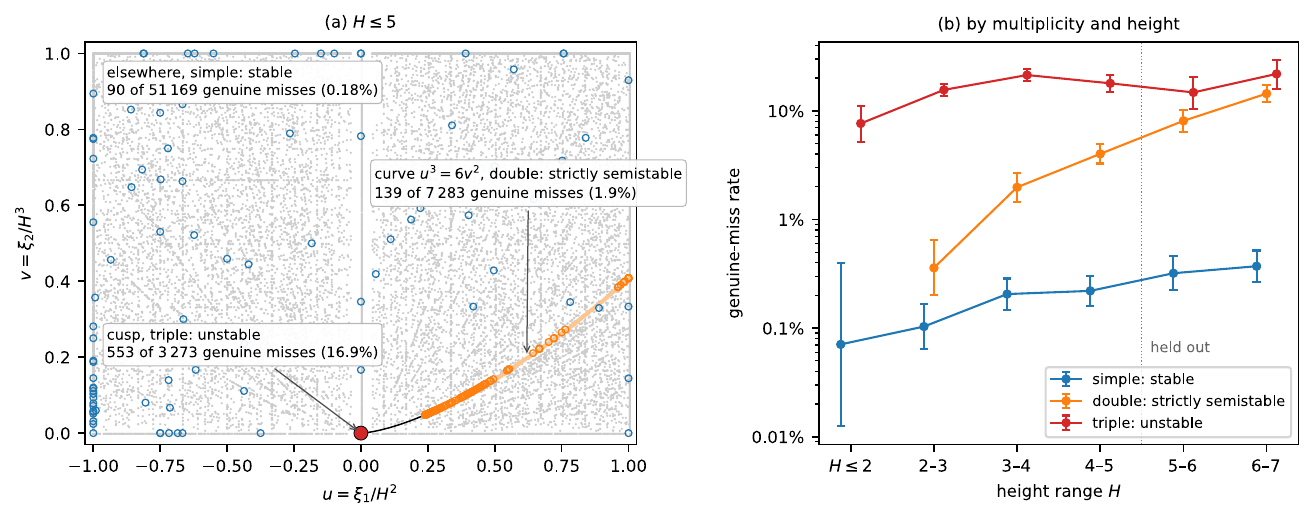}}{\fbox{\parbox[c][5cm][c]{0.9\textwidth}{\centering figure: \texttt{fig\_multiplicity.pdf}}}}
\caption{The hard instances and the multiplicity of the fixed points. (a) The points with $H\le5$ in the plane $(u,v)=(\xi_1/H^2,\xi_2/H^3)$:
double fixed points lie on the curve $u^3=6v^2$, triple ones at its cusp; open circles are genuine misses. (b) The genuine-miss rate by
multiplicity in each shell, with $95\%$ Wilson intervals (logarithmic scale); the shell $H\le2$ has no double fixed point, which needs
$H\ge\sqrt6$.}
\label{fig:multiplicity}
\end{figure}

The genuine-miss rate rises with the multiplicity of the fixed point in every sample and every shell (\cref{tab:multiplicity,fig:multiplicity}):
$0.18\%$, $1.9\%$ and $16.9\%$ below height $5$ for simple, double and triple fixed points. The locus holds $692$ of the $782$ genuine misses ($88\%$) on
$17\%$ of the points of the sample, and, weighted by the census, $84\%$ of them ($48\%$ on the surface, $36\%$ at double fixed points) on $13.9\%$ of the
points; at $5<H\le6$ and $6<H\le7$ it holds $75\%$ and $80\%$ of them on $9.4\%$ and $8.4\%$ of the points. Along the height the rate at a double fixed
point rises steadily, from $0.36\%$ at $2<H\le3$ to $4.0\%$ at $4<H\le5$, $8.1\%$ at $5<H\le6$ and $14.4\%$ at $6<H\le7$, close to the rate on the surface,
while the share of the points on the surface falls faster than the share at double fixed points. The misses therefore move from the triple fixed points
to the double ones (\cref{fig:multiplicity}b): at $6<H\le7$ the double fixed points hold $99$ genuine misses and the surface $33$. Away from the locus the rate stays between
$0.07\%$ and $0.37\%$. The rise holds within each of the two families of double fixed points: for $m=1$ the rate is $0.36\%$, $2.0\%$, $6.8\%$,
$12.8\%$ and $17.6\%$ in the shells from $2<H\le3$ to $6<H\le7$, and for $m=2$ it is $1.6\%$, $3.4\%$ and $11.2\%$ at $2\sqrt6<H\le5$, $5<H\le6$ and
$6<H\le7$. In the terms of invariant theory, the miss rate thus rises from the stable fixed-point quartics to the strictly
semistable ones and to the unstable ones; this describes where the hard instances are, and does not by itself explain them. The locus of double fixed points accounts for the halo of \cref{sec:surfheight}, and for the observation of \cref{sec:general} that the misses move off the
surface as the height grows. The multiplicity class alone ranks the genuine misses with AUC $0.90$ below height $5$ and $0.84$ and $0.86$ on the
held-out ranges, where the classifier, which uses much more, reaches $0.94$ and $0.88$.

\subsection{The surface of triple fixed points}
\label{sec:triple}
The genuine misses concentrate most strongly on the surface $\xi_1=\xi_2=0$ (\cref{tab:nullcone}): its $3\,273$ points,
$5.3\%$ of the sample, contain $553$ of the $782$ genuine misses ($71\%$), a rate of $16.9\%$; on the rest of the moduli space the rate is $0.39\%$,
and neither condition alone matters ($0.12\%$ with $\xi_1=0$ only, $0.31\%$ with $\xi_2=0$ only). By the invariant at which the moduli height is
attained, the rate is $0.10\%$ and $0.29\%$ when it is $\xi_2$ or $\xi_1$, and $2.0\%$ and $4.3\%$ when it is the mixed invariant $\xi_3$ or $\xi_4$,
consistent with the classifier's reliance on the weight-zero coordinates. The invariants $\xi_1=(I,I)_4$ and $\xi_2=((I,I)_2,I)_4$ are the two classical invariants of the fixed-point quartic, of degrees $2$ and $3$
\citep[Eq.~(1)]{m3}, so the surface consists of the maps with a fixed point of multiplicity at least three (\cref{sec:double}): a fixed point of
multiplier $1$ at which three fixed points have merged. As a check of the conventions, on $4\,161$ maps, $1\,226$ of them with a triple fixed point, the
condition $\xi_1=\xi_2=0$ and a root of multiplicity at least three of $I$, computed from the map and factored exactly, agree in every case.

\begin{figure}[ht]
\centering
\begin{minipage}[b]{0.45\textwidth}\centering
\includegraphics[width=\textwidth]{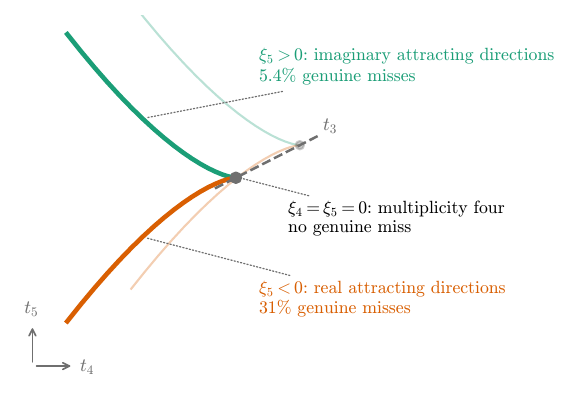}\\[2pt]
\small (a)
\end{minipage}\hfill
\begin{minipage}[b]{0.53\textwidth}\centering
\IfFileExists{fig_surface3d.pdf}{\includegraphics[width=\textwidth]{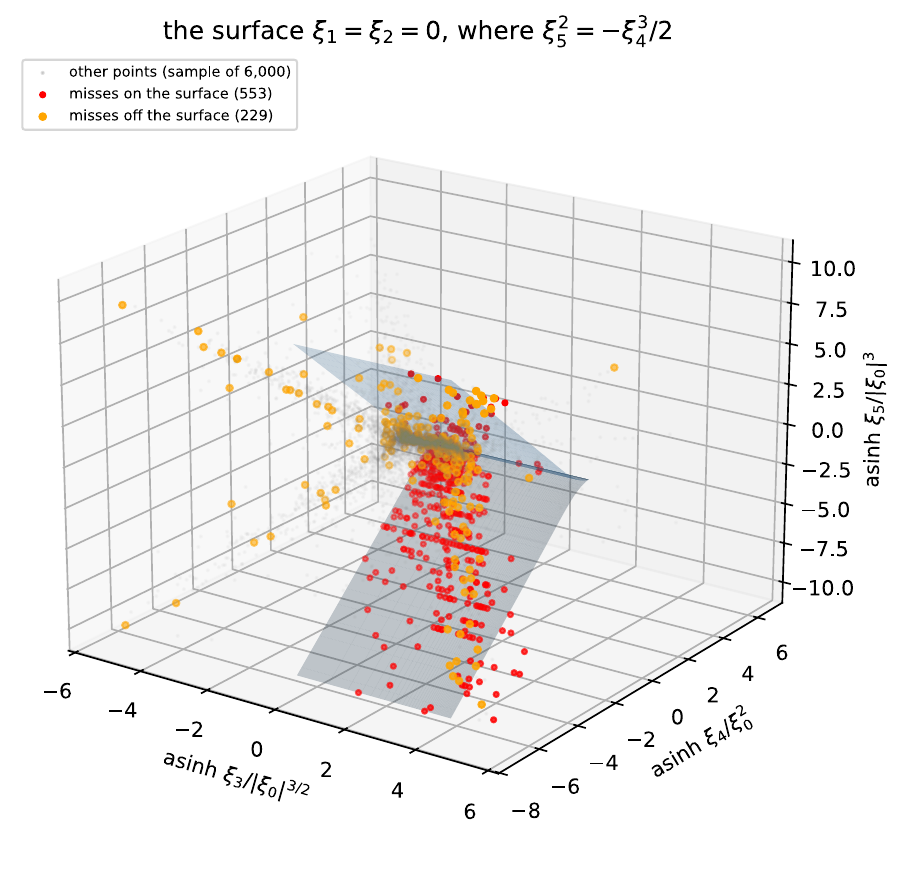}}{\fbox{\parbox[c][5cm][c]{0.9\textwidth}{\centering figure: \texttt{fig\_surface3d.pdf}}}}\\[2pt]
\small (b)
\end{minipage}
\caption{The surface $\xi_1=\xi_2=0$. (a) Schematic cross-section at fixed $t_3$, not to scale: the cusp $t_5^2=-\frac12t_4^3$, with
the branch $\xi_5>0$ (imaginary attracting directions), the branch $\xi_5<0$ (real ones) and the cusp line, where the fixed point has
multiplicity four; rates from the sample with $H\le5$. (b) The data in the coordinates $t_i=\xi_i/|\xi_0|^{w_i/2}$, on $\operatorname{asinh}$
axes: the surface (blue), the genuine misses on it (red) and off it (orange), and $6\,000$ other points (grey).}
\label{fig:surface}
\end{figure}

On the surface the relation \citep[Eq.~(2)]{m3}, which is the relation \citep[Thm.~3.2]{west} in other notation, reduces to
\[
\xi_5^2=-\tfrac12\xi_4^3.
\]
Hence $\xi_4\le0$, and for integer coordinates
$\xi_4=-2u^2$ and $\xi_5=2u^3$ with $u\in\Z$ of the sign of $\xi_5$, since $-\tfrac12\xi_4^3$ is a square only if $\xi_4=-2k$ with $k$ a square: the
integer points of the surface are
\[
(\xi_0,\xi_1,\dots,\xi_5)=(\xi_0,0,0,\xi_3,-2u^2,2u^3),\qquad u\in\Z,
\]
and all $221\,178$ points of the surface in the census to $H\le5$ have this form.
In the coordinates $t_i=\xi_i/|\xi_0|^{w_i/2}$ the surface is the cuspidal cylinder $t_5^2=-\tfrac12t_4^3$, swept along $t_3$ (\cref{fig:surface}).
An example is the map $\phi(z)=(z^3+z+1)/z^2$ \citep[Ex.~8.1]{m3}, with moduli point $[8:0:0:16:-2:-2]$, whose fixed points are $-1$ and a triple
fixed point at $\infty$. All points of the surface are fine: on it $-m_{12}=\xi_3^2+2\xi_0u^2$ is a norm from $\Q(\sqrt{-2\xi_0})$, so the symbol $(-m_{12},-2\xi_0)$ \citep[Thm.~3.3]{m3} is trivial (equivalently, the unique triple fixed point gives a rational point on the descent conic). The
parametrisation explains why the surface thins out with the height: it has about $B^7$ points of height at most $B$ ($|\xi_0|\le B^2$,
$|\xi_3|\le B^3$, $|u|\ll B^2$), against the heuristic prediction $N(B)\asymp B^8\log B$ for all of $\M_3(\Q)$ \citep[\S4]{m3}, which suggests that its share decreases roughly
like $1/(B\log B)$.
Among all points of the census (coarse points and points with automorphisms included), its share is $9.9\%$, $4.6\%$, $3.1\%$ and $1.8\%$ of the
points of height at most $2,3,4,5$, roughly consistent with $1/(B\log B)$, over a range too short to estimate an exponent.

\begin{table}[ht]
\centering
\caption{Genuine misses by position of the moduli point ($61\,725$ points, $782$ genuine misses, rate $1.27\%$). Left: on and off
the surfaces $\xi_1=0$ and $\xi_2=0$. Right: by the invariant at which the moduli height is attained.}
\label{tab:nullcone}
\small
\begin{adjustbox}{max width=\textwidth}
\begin{tabular}{lrrr@{\qquad}lrr}
\toprule
 & points & misses & rate & height attained at & points & rate\\
\midrule
$\xi_1=\xi_2=0$ & 3\,273 & 553 & 16.9\% & $\xi_0$ & 15\,371 & 1.09\%\\
$\xi_1=0\neq\xi_2$ & 4\,107 & 5 & 0.12\% & $\xi_1$ & 13\,425 & 0.29\%\\
$\xi_2=0\neq\xi_1$ & 2\,552 & 8 & 0.31\% & $\xi_2$ & 12\,612 & 0.10\%\\
$\xi_1\xi_2\neq0$ & 51\,793 & 216 & 0.42\% & $\xi_3$ & 13\,106 & 1.95\%\\
 & & & & $\xi_4$ & 7\,178 & 4.28\%\\
 & & & & $\xi_5$ & 33 & 0\%\\
\bottomrule
\end{tabular}
\end{adjustbox}
\end{table}

\subsection{The two branches and their dynamics}
\label{sec:halves}
\begin{figure}[t]
\centering
\IfFileExists{fig_surface_halves.pdf}{\includegraphics[width=0.6\textwidth]{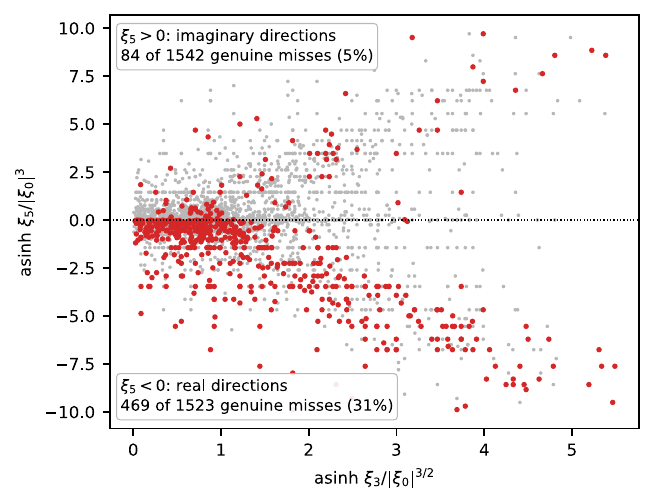}}{\fbox{\parbox[c][5cm][c]{0.55\textwidth}{\centering figure: \texttt{fig\_surface\_halves.pdf}}}}
\caption{The two halves of the surface, for its points with $H\le5$: position along the cylinder ($\operatorname{asinh}t_3$) against
$\operatorname{asinh}t_5$, whose sign separates the halves (dotted, the cusp line). Red, the genuine misses.}
\label{fig:halves}
\end{figure}

 The moduli space is a double cover of the weighted space of the first five invariants, and the two points over the same
$(\xi_0,\dots,\xi_4)$ are $(\xi_0,\dots,\xi_4,\pm\xi_5)$ \citep[\S2.1]{m3}. On the surface they form its two halves $\xi_5<0$ and $\xi_5>0$, which turn out
to be hard at very different rates; we give the sign of $\xi_5$ a dynamical meaning. Where $\xi_3\neq0$, that is, on the surface away from the stratum $L_1=\{\xi_2=\xi_3=0\}$ of maps with an involution
\citep{jsc}, and so at every generic point, the sign of $\xi_5$ is an invariant of the point: two rational representatives then differ by $\xi_i\mapsto\lambda^{w_i}\xi_i$
with $\lambda\in\Q^\times$ \citep[\S2.2]{m3}, and $\lambda^6>0$. They behave very differently (\cref{fig:halves}). On the half $\xi_5<0$ ($1\,523$
points) the genuine-miss rate is $30.8\%$; on the half $\xi_5>0$ ($1\,542$ points) it is $5.4\%$; on the cusp line $\xi_4=\xi_5=0$ ($208$ points) there
is no genuine miss. The half $\xi_5<0$, $2.5\%$ of the sample, contains $469$ of the $782$ genuine misses ($60\%$). The asymmetry does not run through
the size of the resultant: on the surface the resultant invariant is $I_6=\tfrac1{256}\xi_3^2-\tfrac1{64}\xi_0\xi_4+\tfrac18\xi_5$, and among points
with $|I_6|$ in the same fifth of its range, from the smallest to the largest, the rate on the half $\xi_5<0$ is $15\%$, $20\%$, $39\%$, $37\%$ and
$43\%$, and on the half $\xi_5>0$ it is $0.6\%$, $1.4\%$, $4.1\%$, $4.4\%$ and $14\%$ (about $300$ points in each cell; the smallest rate is two
misses).

The sign of $\xi_5$ has a dynamical meaning, found by the same move as the surface itself: the figure showed the misses
on one wing, and we asked what distinguishes that wing. At a triple fixed point $p$ the map has the local form
\[
z\mapsto z+a(z-p)^3+\dots
\]
in any rational coordinate in which $p$ is finite, and the sign of $a$ is an invariant of the rational class: a rational change of coordinates
$z=\gamma(w)$ with $\gamma(q)=p$ and $q$ finite multiplies $a$ by $\gamma'(q)^2$, the square of a nonzero rational number.

\begin{proposition}\label{prop:sign}
Let $\phi$ be a cubic map over $\Q$ with a fixed point of multiplicity at least three and $\xi_3\neq0$, and $a$ the cubic coefficient there, in
any rational coordinate in which the fixed point is finite. Then
$\operatorname{sign}(a)=\operatorname{sign}(\xi_5)$, and $a=0$ exactly on the line $\xi_4=\xi_5=0$, where the fixed point has multiplicity four.
\end{proposition}

\begin{proof}
A root of multiplicity at least three of a quartic is unique, hence fixed by Galois, so the triple fixed point is rational and a rational
change of coordinates moves it to $\infty$. The coefficients of $x^4$, $x^3y$ and $x^2y^2$ in $I=yf_0-xf_1$ are $-c_4$, $c_0-c_5$
and $c_1-c_6$, so requiring $y^3\mid I$ forces $c_4=0$, $c_5=c_0$, $c_6=c_1$, so that
\[
\phi(z)=\frac{c_0z^3+c_1z^2+c_2z+c_3}{c_0z^2+c_1z+c_7},\qquad c_0\neq0,\qquad I=y^3\bigl((c_2-c_7)x+c_3y\bigr).
\]
For these maps the invariants \citep[Eq.~(1)]{m3} are $\xi_1=\xi_2=0$, $\xi_4=-2c_0^2(c_2-c_7)^2$ and $\xi_5=-2c_0^3(c_2-c_7)^3$, while in the coordinate $w=1/z$ the map is
$w\mapsto w+aw^3+\dots$ with $a=(c_7-c_2)/c_0$. Hence
\[
\xi_5=2c_0^6\,a^3 ,
\]
and since a change of coordinates multiplies $a$ by a square and, because $\xi_3\neq0$ and $\xi_4=-2u^2\neq0$ whenever $\xi_5\neq0$, the
weights of the nonzero coordinates have greatest common divisor $1$, so that two rational representatives of the point differ by
$\xi_i\mapsto\lambda^{w_i}\xi_i$ \citep[\S2.2]{m3}, a sixth power on $\xi_5$, the signs agree. If $a=0$, then $c_2=c_7$, $\xi_4=\xi_5=0$ and $I=c_3y^4$, with $c_3\neq0$ since the resultant is nonzero.
\end{proof}

The hypothesis $\xi_3\neq0$ is needed. When $\xi_2=\xi_3=0$ as well, rational representatives may differ by $\xi_i\mapsto\mu^{w_i/2}\xi_i$ with
$\mu\in\Q^\times$, which changes the sign of $\xi_5$ when $\mu<0$: the maps $z^3/(z^2-1)$ and $z^3/(z^2+1)$, both with a triple fixed point at
$\infty$, have $a=-1$ and $a=1$ but the same moduli point $[24:0:0:0:-2:2]$. Such points lie in $L_1$: both maps commute with $z\mapsto-z$. They are outside our samples, which contain generic points only.

The identity checks on $40$ maps from the census, $20$ from each half, and the proof recovers the parametrisation of the surface with
$u=-c_0(c_2-c_7)$. For $w\mapsto w+aw^3+\dots$ the attracting directions are the $v$ with $av^2<0$ \citep[\S10]{milnor-book}, so the two attracting directions of the parabolic fixed point
are real for $a<0$ and imaginary for $a>0$: at small heights, the hard instances with a triple
fixed point are mainly those whose attracting directions are real.

Shell by shell, the genuine-miss rate on the half $\xi_5<0$ is $18\%$, $31\%$, $34\%$, $29\%$, $25\%$ and $24\%$ in the
shells $H\le2$ to $6<H\le7$, and on the half $\xi_5>0$ it is $0.8\%$, $2.6\%$, $9.0\%$, $9.0\%$, $3.6\%$ and $21\%$. In every shell up to $5<H\le6$
the $95\%$ Wilson intervals of the two halves are disjoint (at $5<H\le6$, $[17\%,34\%]$ against $[1.2\%,10\%]$, on $24$ of $97$ and $3$ of $84$
points); at $6<H\le7$ they overlap ($[15\%,36\%]$ and $[14\%,30\%]$, on $15$ of $63$ and $18$ of $87$). The asymmetry is therefore clear up to
height $6$; whether it fades beyond, as the last shell suggests, one shell cannot tell. The surface as a whole remains where the hard instances are.

\subsection{How the locus was found}
\label{sec:found}
It was the classifier, through its reliance on $\xi_1/H^2$ and $\xi_2/H^3$, that pointed to the surface of triple fixed points.
Because the surface is a coordinate locus, a cross-tabulation of the misses against the vanishing of each $\xi_i$ and of each pair would also have found it; for a locus that is not a coordinate hyperplane, asking what a fitted model relies on is the more direct route. The model suggested that small values of $\xi_1/H^2$ and $\xi_2/H^3$ matter, and only the integers showed that the condition is exact
vanishing: of the $3\,452$ points with $\xi_2=0$ and $\xi_1/H^2$ in its lowest fifth, the $3\,273$ with $\xi_1=0$ carry all $553$ genuine misses and
the other $179$ none. On integer data a correlation with a continuous normalisation can disguise an exact algebraic condition, and the two have
different consequences.

Unlike the surface, the hypersurface was suggested by theory rather than by the classifier. Reduction theory \citep{hutz-stoll} applies to stable binary forms, and the fixed-point quartic fails to be stable not only on the surface but on the whole
hypersurface: it is strictly semistable at a double fixed point and unstable at a triple or quadruple one. We therefore tested the hypersurface as
a whole, after the analyses of the surface and after the held-out samples of \cref{sec:general} had been evaluated. In hindsight the classifier had already
pointed to it: the halo it finds around the surface (\cref{sec:surfheight}) is precisely the locus of double fixed points.

\subsection{The locus and the height}
\label{sec:surfheight}
 The surface becomes more dangerous and rarer as the height grows, and rarer faster
(\cref{tab:shellsurface}). Its share of the sample, which consists of fine generic points with $\xi_0m_{12}\neq0$ and counts each shell separately, falls from $18\%$ at
$H\le2$ to $2.7\%$ at $4<H\le5$, as the count above predicts, while
the miss rate on it rises from $8\%$ to $16$--$21\%$; at small heights $\halg$ is small and there is little room for a poorly reduced model to be far from
minimal. Away from the surface the miss rate grows from $0.07\%$ to $0.65\%$. In the outermost shell the misses off the surface already outnumber
those on it ($0.63\%$ against $0.49\%$ of all points), and the flat overall rate of \cref{tab:misses} is the sum of these opposite trends. Most of the
misses off the surface there are at a double fixed point ($88$ of $126$; \cref{sec:double}).

Off the surface the classifier still finds the misses, and not because they have larger answers. On the $58\,452$ points off it ($229$ genuine
misses, base rate $0.39\%$) the classifier gives AUC $0.85$ $[0.82,0.88]$ and AP $0.084$, a $22$-fold enrichment, against AUC $0.68$ for the size of the algorithm's
answer. Its capture curve is steep and then flat: its top $5\%$ contains $59$--$60\%$ of the off-surface
misses and its next $5\%$ only $6$--$9\%$ more, so off the surface the triage budget should not exceed about $5\%$. Part of what it finds is a halo around the
surface: among the off-surface points, the tenth with the smallest $\max(|\xi_1|^{1/2},|\xi_2|^{1/3})/H$ has a miss rate of $1.6\%$, $4.1$ times
the off-surface average, and holds $93$ of the $229$ misses. The halo is the locus of double fixed points of \cref{sec:double}: without those points the same group
has a rate of $0.05\%$, and the rate off the surface is then lowest next to it rather than highest.

\begin{table}[ht]
\centering
\caption{The surface by shell, and the surface rule: the deep search on the surface and on the top $5\%$ (or $10\%$) of the other points by a
classifier trained off the surface.}
\label{tab:shellsurface}
\begin{adjustbox}{max width=\textwidth}
\begin{tabular}{lcccc}
\toprule
shell & $H\le2$ & $2<H\le3$ & $3<H\le4$ & $4<H\le5$\\
\midrule
share of the points on the surface & 18.2\% & 7.3\% & 4.8\% & 2.7\%\\
genuine-miss rate on the surface & 7.6\% & 15.6\% & 21.3\% & 17.9\%\\
genuine-miss rate off the surface & 0.07\% & 0.15\% & 0.39\% & 0.65\%\\
genuine misses on the surface & 96\% & 89\% & 73\% & 43\%\\
\midrule
surface rule: points searched & 18.6\% & 8.6\% & 10.4\% & 10.4\%\\
surface rule: genuine misses found & 96\% & 93\% & 87\% & 84\%\\
\midrule
with $10\%$ off the surface: points searched & 19.4\% & 11.0\% & 16.0\% & 16.8\%\\
with $10\%$ off the surface: genuine misses found & 96\% & 93\% & 89\% & 87\%\\
\bottomrule
\end{tabular}
\end{adjustbox}
\end{table}

\section{Search reach versus reduction}
\label{sec:why}

Two explanations are possible for the concentration of the hard instances on the locus: the reduction step of Algorithm~A, which loses one of its
covariants there, and the reach of its local search. The data favour the second.

\subsection{The reduction step}
A natural suspect is the reduction at infinity of step 4. One of its covariant points is built from the roots $\alpha_j$ of $I$: the minimiser of
$\sum_j t_j|z-\alpha_j|^2/\operatorname{Im}z$ with inverse-derivative weights $t_j=|I'(\alpha_j)|^{-1}$ \citep[Alg.~7.1(4)]{m3}. These weights are
undefined at a multiple root, that is, on the whole locus of \cref{sec:double} and not only on the surface, and there Algorithm~A does not use this
point (\cref{app:algA}): it reduces with the covariant points of the quadratics alone. This construction is not the covariant point of reduction theory \citep[\S3]{hutz-stoll}, the minimiser of a function of the roots defined
for stable forms, so the step-4 figures below describe step 4 of Algorithm~A, not reduction at infinity in that sense. The data support only
half of the suspicion. Step 4 is degraded on the surface: step 5 has to lower the height left by step 4 at $41\%$ of the points of the half
$\xi_5<0$, $64\%$ of the half $\xi_5>0$ and $59\%$ of the cusp line, against $11\%$ off the surface. But the measured degradation does not follow the asymmetry of the miss rates
between the two halves: it is largest where Algorithm~A rarely misses, on the cusp line, where the model after step 4 is twice the size of the best model at the median and at least twice at
$54\%$ of the points, and on the half $\xi_5>0$ (median ratio $1.13$); there Algorithm~A rarely or never misses, because step 5 repairs it. On the
half $\xi_5<0$, where the misses are, the model after step 4 is almost as close to the best in height as off the surface (median ratio $1.04$ against
$1.00$, at least twice the best at $1.8\%$ of the points against $2.3\%$), yet the smaller model lies four or five moves away. The evidence therefore
favours an explanation in terms of reachability by the local search, without ruling out an interaction with the reduction at infinity;
\cref{sec:welldefined} measures that interaction directly, and finds it small.

\subsection{The paths to the smaller models}
\label{sec:paths}
The recorded paths say what is out of reach. On the half $\xi_5<0$, $64\%$ of the genuine misses need depth $5$, against $31\%$ on the half
$\xi_5>0$ and $15\%$ off the surface, and the paths share a recurring shape: $344$ of the $469$ ($73\%$) fall into eight sets of moves, and seven of
these, with $328$ paths, consist of exactly one inversion $S\colon z\mapsto-1/z$, exactly one translation $T\colon z\mapsto z+1$ and two or three
conjugations of determinant $2$ or $3$; over all $469$ paths there are on average $1.12$ inversions, $0.99$ translations and $2.45$ such
conjugations, and the $p$-adic moves come first in $314$ of the $469$ paths ($117$ interleave the two kinds, $38$ begin with the unimodular moves). On
the half $\xi_5>0$ the order is reversed: $83$ of the $84$ paths begin with the unimodular moves. Off the surface the three orders are mixed ($48$,
$64$ and $116$). Primes above $3$ hardly occur on the surface ($15$ of the $469$ paths on the half $\xi_5<0$, none of the $84$ on the half
$\xi_5>0$, against $32$ of the $229$ off the surface): for all but $15$ of the $469$ paths on the half $\xi_5<0$, step 5 has the moves and lacks only
depth. The unimodular part $ST$ has trace $1$ and $(ST)^3=-I$: it is the elliptic element of order three of $\mathrm{PSL}_2(\Z)$ fixing
$\rho=e^{2\pi i/3}$. On
the half $\xi_5>0$ the misses rarely use the translation ($0.17$ per path against $0.99$). The data do not show why the half with real attracting directions needs the elliptic correction and the other half does not.

At the double fixed points the improvements also lie just beyond the radius of the local search, with
improving paths of $4.1$ to $4.3$ moves on average, and they also move only at the primes $2$ and $3$ ($99$--$100\%$ of the paths, against at most
$2\%$ at a prime above $3$). But only $10$--$30\%$ of these paths contain the elliptic pattern, an inversion next to a translation, against $45$--$72\%$
on the surface: the recurring shape above belongs to the triple fixed points. The targeted repair of \cref{sec:method}, read off the paths on the surface,
still recovers $69\%$, $69\%$ and $74\%$ of the genuine misses at double fixed points in the three samples ($96$ of $139$, $42$ of $61$, $73$ of $99$),
against $43$--$65\%$ of those away from the locus. The model after step 4 is slightly further from Algorithm~A's answer at a double fixed point (median
ratio of heights $1.06$, against $1.00$ at simple fixed points and $1.02$ on the surface).

\subsection{A well-defined reduction}
\label{sec:welldefined}
The hard instances and the gap in the reduction of \cref{sec:why}, where step 4 reduces with the covariant points of the quadratics alone, occupy
the same locus, so the hard instances could be an artefact of reducing with less there. We
tested this directly. On every point of the locus in the three samples ($10\,556$, $938$ and $839$ points) we ran Algorithm~A with one covariant
point added to step 4: either the point built from the roots of $I$, computed with $|I'(\alpha_j)|$ bounded below by $10^{-12}$ at the repeated root, or the covariant point of the dynatomic form of period two,
\[
\Phi_2^*=\frac{yF_2-xG_2}{yf_0-xf_1},\qquad \phi\circ\phi=[F_2:G_2],
\]
a form of degree $6$, which reduction theory uses when the fixed-point form is not stable \citep[\S6]{hutz-stoll}. A fixed point is a root of $\Phi_2^*$ only when its multiplier is $-1$ \citep{silverman-ads}, so $\Phi_2^*$ is not degenerate at a fixed
point of multiplier $1$. It was stable at all but one of the $12\,333$ points: at $[24:6:6:-16:-48:-64]$, a double fixed point with $m=1$, where $\infty$ is a fixed point of
multiplier $-1$, all six points of period two coincide with it, $\Phi_2^*$ is a multiple of $y^6$, and no point was added. Everything else in Algorithm~A was unchanged, and without the added
point the same program reproduces Algorithm~A's stored answer at every point. The covariant point of $\Phi_2^*$ removes $121$ of the $726$ misses
on the locus below height $5$ and creates $13$, and removes $24$ of $98$ and $26$ of $145$ at $5<H\le6$ and $6<H\le7$, creating $4$ and $6$
(\cref{tab:step4}); the bounded point removes $82$, $13$ and $16$ and creates $35$, $3$ and $10$. The well-defined point helps at double fixed points
($74$ of $321$ misses over the three samples, $23\%$) and on the surface ($97$ of $648$, $15\%$), and on a random control sample of $2\,000$ points
it changes nothing away from the locus: the three misses it removes there all lie on it. About five hard instances in six therefore survive a reduction that uses a covariant defined where they
occur: the reduction step contributes, but the hard instances appear to be mainly a matter of the reach of the local search, as the paths of \cref{sec:paths}
suggest. Why the smaller models of maps with a multiple fixed point lie beyond that reach is a question our experiments do not answer.

\begin{table}[ht]
\centering
\caption{Misses of Algorithm~A on the locus of multiple fixed points, and on $2\,000$ random points, with one covariant point added to step 4:
the point built from the roots of $I$ with bounded weights, or the covariant point of $\Phi_2^*$. In parentheses: misses removed / created.}
\label{tab:step4}
\small\setlength{\tabcolsep}{4pt}
\begin{adjustbox}{max width=\textwidth}
\begin{tabular}{llrrcc}
\toprule
sample & fixed points & points & as published & bounded point & $\Phi_2^*$ covariant\\
\midrule
$H\le5$ & double & $7\,283$ & $149$ & $150$ ($31/32$) & $126$ ($32/9$)\\
 & triple & $3\,273$ & $577$ & $529$ ($51/3$) & $492$ ($89/4$)\\
$5<H\le6$ & double & $755$ & $67$ & $58$ ($12/3$) & $52$ ($19/4$)\\
 & triple & $183$ & $31$ & $30$ ($1/0$) & $26$ ($5/0$)\\
$6<H\le7$ & double & $688$ & $105$ & $102$ ($13/10$) & $87$ ($23/5$)\\
 & triple & $151$ & $40$ & $37$ ($3/0$) & $38$ ($3/1$)\\
\midrule
$H\le5$, random & all & $2\,000$ & $20$ & $19$ ($1/0$) & $17$ ($3/0$)\\
\bottomrule
\end{tabular}
\end{adjustbox}
\end{table}

\section{A certified structured search: Algorithm A$^+$}
\label{sec:method}

Algorithm~A$^+$ adds two components to Algorithm~A, each motivated by \cref{sec:misses,sec:surface,sec:why}, and returns the smallest verified model
it finds, Algorithm~A's own answer included.

\subsection{Targeted repair}

The shape of the missing improvements (\cref{sec:paths}) suggests a repair. From the model after step 4 we try every
chain of zero to three conjugations at $2$ and $3$, by $\left(\begin{smallmatrix}p&a\\0&1\end{smallmatrix}\right)$ and
$\left(\begin{smallmatrix}1&0\\0&p\end{smallmatrix}\right)$, followed by every word of length at most two in $z\pm1$, $-1/z$, $-z$, $(1+7+7^2+7^3)(1+4+4^2)=8\,400$
candidates, and from the best of them run a local search with step 5's eleven moves. This final search is the
first-improvement search of \cref{sec:checks}, which costs the same as step 5; since Algorithm~A$^+$ keeps the smallest verified candidate, the choice
affects only how much the repair finds, not the correctness of the result. Every repaired model is verified by recomputing its invariants, and a model
that fails verification is rejected. The move set was read off the paths of the genuine misses on the half $\xi_5<0$ with $H\le5$, so the
result there is partly in-sample.

\Cref{tab:targeted} shows that the repair is effective also beyond the points used to design it. It recovers $59\%$ of the genuine misses off the
surface, which played no part in its design, and about two thirds of the genuine misses in both held-out height ranges; it is weakest on the small
upper half of the surface, and on misses that need a prime above $3$, which it does not use; in the held-out ranges it therefore fixes only about
$60\%$ of all misses. This suggests that the composite moves capture a recurring limitation of the original search, not a feature of the development set. Structure appears to matter more than radius: the reference search stopped at depth $4$, which moves at more primes than the repair and whose
depth-$4$ ball has up to about $16\,000$ models even where it moves only at $2$ and $3$, reaches $424$ of the $782$ genuine misses at $H\le5$,
against $544$ for the repair with $8\,400$ candidates. This compares what the two searches find, not what they cost: the repair's candidates are
followed by a local search of step 5's cost, so the two are not a controlled comparison of work. The repair needs no classifier, runs on every point, and costs about $0.25$\,s beyond Algorithm~A's own steps
(\cref{sec:runtime}). On $91$ of the $3\,273$ points of the surface, and on $24$ and $38$ points of the held-out samples, it alone returns a larger model than
Algorithm~A, which costs nothing since Algorithm~A$^+$ keeps the smallest candidate. On two points it finds a model smaller than both Algorithm~A and the reference search: $[23:0:0:22:-32:-128]$ at $H\le5$ (repair $89$, Algorithm~A $113$, reference $107$) and $[-23:0:0:168:-512:-8192]$ at
$5<H\le6$ (repair $341$, Algorithm~A and reference $361$).

\begin{table}[ht]
\centering
\caption{The targeted repair, from the model after step 4. Recovered: the repaired model is at least as small as the reference search's best.
Role: development (the move set was read off these paths), not used (same heights), held out (larger heights).}
\label{tab:targeted}
\small\setlength{\tabcolsep}{4pt}
\begin{adjustbox}{max width=\textwidth}
\begin{tabular}{llcc}
\toprule
points & role & genuine misses & all misses\\
 & & recovered & fixed\\
\midrule
surface, $H\le5$, half $\xi_5<0$ & development & $377/469$ ($80\%$) & \\
surface, $H\le5$, half $\xi_5>0$ & not used & $32/84$ ($38\%$) & \\
surface, $H\le5$, all $3\,273$ points & mixed & $409/553$ ($74\%$) & $410/577$ ($71\%$)\\
off the surface, $H\le5$ (genuine misses) & not used & $135/229$ ($59\%$) & \\
all misses, $H\le5$ & mixed & $544/782$ ($70\%$) & $545/836$ ($65\%$)\\
random sample, $H\le5$ ($2\,000$ points) & mostly not used & $13/19$ & $13/20$\\
$5<H\le6$, all $10\,000$ points & held out & $76/117$ ($65\%$) & $76/130$ ($58\%$)\\
$6<H\le7$, all $10\,000$ points & held out & $110/166$ ($66\%$) & $110/188$ ($59\%$)\\
\bottomrule
\end{tabular}
\end{adjustbox}
\end{table}

\subsection{Triage}
\label{sec:triagecls}
The depth-$5$ reference search is too expensive to run everywhere: it examines about $236\,000$ models per point on average, most of them at a
minority of points with several small primes in the resultant, and takes about eleven times the running time of Algorithm~A (\cref{sec:runtime}). A gradient-boosted classifier, trained on the genuine-miss labels with the features D and $\log\halg$ (\cref{sec:features}), scores each point
after Algorithm~A has run, and
the deep search is run only where the score reaches a fixed threshold $t$. Both are fixed on the points with $H\le5$: $t=0.4824$ is the $95$th
percentile of the out-of-fold scores there, each point being scored by a model trained on the other four fifths, and for new points the classifier
is refitted on all points with $H\le5$ while $t$ is kept. This rule, fixed before the held-out samples were evaluated, defines Algorithm~A$^+$ in
\cref{sec:eval}.

A second rule needs no learning: run the deep search on the maps with a multiple fixed point, $\xi_1^3=6\xi_2^2$ (\cref{sec:double}). It has no fitted
parameter, but it was found after the held-out samples had been evaluated, so we report it as a variant, Algorithm~A$^+$ with the exact rule.

\subsection{The algorithm}
\begin{algorithm}[H]
\caption{Algorithm~A$^+$}\label{alg:Aplus}
\begin{algorithmic}[1]
\Require a fine generic moduli point $P$ with $\xi_0m_{12}\neq0$; a selection rule of \cref{sec:triagecls}
\State $\phi_A\gets$ Algorithm~A$(P)$, and $\phi_4\gets$ its model after step 4 \Comment{\citep[Alg.~7.1]{m3}}
\State $\mathcal C\gets\{\phi_A\}$
\State $\mathcal C\gets\mathcal C\cup\{\textsc{TargetedRepair}(\phi_4)\}$ \Comment{$8\,400$ composite moves, then a local search}
\If{the rule selects $P$} \Comment{$f(P,\log h(\phi_A))\ge t$, or $\xi_1^3=6\xi_2^2$}
  \State $\mathcal C\gets\mathcal C\cup\{\textsc{ReferenceSearch}(\phi_4)\}$ \Comment{depth $5$, primes up to $13$}
\EndIf
\State \Return a model of least height among those in $\mathcal C$ whose invariants agree with $P$
\end{algorithmic}
\end{algorithm}

\begin{remark}\label{prop:nonworse}
The model $\phi^+$ returned by Algorithm~A$^+$ lies in the class $P$, and $\hmin(P)\le h(\phi^+)\le\halg(P)$.
\end{remark}

\begin{proof}
Every candidate is accepted only if its invariants agree with $P$ up to weighted scaling, so $\phi^+$ lies in the class $P$ and
$h(\phi^+)\ge\hmin(P)$ by definition; and $\phi^+$ is the smallest of a set of models that contains Algorithm~A's answer.
\end{proof}

The remark is immediate, since Algorithm~A's own answer is one of the candidates. Its point is that the correctness of Algorithm~A$^+$ does not
depend on the classifier: a wrong decision of the selection rule can only waste time, by searching a point needlessly, or miss an improvement, by
not searching a point that has one. It can never produce a model outside the class or larger than Algorithm~A's, because every candidate is verified
by recomputing its invariants, as our implementation enforces.

\section{Experimental evaluation}
\label{sec:eval}

\subsection{Protocol}
At $H\le5$ Algorithm~A$^+$ is evaluated with the out-of-fold scores of \cref{sec:triagecls}, so that no point is triaged by a classifier that saw
it, and exactly $5\%$ of the points are flagged. On the two held-out ranges the refitted classifier and $t$ are applied unchanged; they flag $5.0\%$
of the points at $5<H\le6$ and $5.1\%$ at $6<H\le7$, the same share as at the training heights. Each component of Algorithm~A$^+$ depends only on the
point, so its result on a point is determined by the stored runs of Algorithm~A, the repair and the reference search on that point, and the
evaluation, computed from those runs, is exact rather than simulated. All relative costs are taken from the per-point benchmark of
\cref{sec:runtime}.

\subsection{Results below height 5, and ablations}
\label{sec:results}
\Cref{tab:aplus} summarises Algorithm~A$^+$ with each selection rule on all three samples. With the classifier rule it reaches the reference
search's best model on $89\%$ of the misses at $H\le5$ and $91\%$ of the genuine ones (these points include those from whose paths the repair was
designed); its two components find largely different misses, and each alone recovers far fewer.

\begin{table}[H]
\centering
\caption{Algorithm~A$^+$ on the three samples: misses / genuine misses of Algorithm~A for which the reference search's best model is
reached, share of the points searched, and time relative to Algorithm~A (measured below height $6$ only). The exact rule was found after
the held-out samples were evaluated.}
\label{tab:aplus}
\small\setlength{\tabcolsep}{4pt}
\begin{adjustbox}{max width=\textwidth}
\begin{tabular}{lccc}
\toprule
 & $H\le5$ & $5<H\le6$ (held out) & $6<H\le7$ (held out)\\
\midrule
misses / genuine misses of Algorithm~A & $836$ / $782$ & $130$ / $117$ & $188$ / $166$\\
\midrule
targeted repair alone & $545$ / $544$ & $76$ / $76$ & $110$ / $110$\\
triage alone (classifier) & $686$ / $659$ & $92$ / $84$ & $102$ / $92$\\
\midrule
Algorithm~A$^+$, classifier rule & $742$ / $715$ ($89\%$ / $91\%$) & $106$ / $98$ ($82\%$ / $84\%$) & $147$ / $137$ ($78\%$ / $83\%$)\\
\quad points searched; time & $5.0\%$; $1.8\times$ & $5.0\%$; $1.6\times$ & $5.1\%$\\
Algorithm~A$^+$, exact rule & $765$ / $731$ ($92\%$ / $93\%$) & $112$ / $102$ ($86\%$ / $87\%$) & $167$ / $154$ ($89\%$ / $93\%$)\\
\quad points searched; time & $17.1\%$; $2.6\times$ & $9.4\%$; $1.5\times$ & $8.4\%$\\
Algorithm~A$^+$, either rule & $766$ / $732$ ($92\%$ / $94\%$) & $113$ / $103$ ($87\%$ / $88\%$) & $168$ / $155$ ($89\%$ / $93\%$)\\
\quad points searched; time & $17.5\%$; $2.6\times$ & $10.0\%$; $1.6\times$ & $9.2\%$\\
\bottomrule
\end{tabular}
\end{adjustbox}
\end{table}

Running the deep search only on the top of the classifier's ranking recovers most of the misses at a small fraction of the cost
(\cref{tab:capture}): the $5\%$ ranked highest cost about $1.3$ times Algorithm~A and contain $83\%$ of all misses and $84\%$ of the genuine ones.

\begin{table}[ht]
\centering
\caption{Triage: misses found when the deep search runs only on the given share of the points ranked highest by the classifier (out-of-fold),
and the estimated mean time per point. The two rows use classifiers trained on the respective labels. Times are estimated from search-node counts, relative to the mean time of Algorithm~A over all
$61\,725$ points ($0.394$\,s); \cref{tab:benchmark} gives direct measurements.}
\label{tab:capture}
\begin{adjustbox}{max width=\textwidth}
\begin{tabular}{lccccc}
\toprule
share of points searched & 2\% & 5\% & 10\% & 20\% & 100\%\\
\midrule
all misses found & 65\% & 83\% & 89\% & 93\% & 100\%\\
genuine misses found & --- & 84\% & 89\% & --- & 100\%\\
estimated mean time per point & 0.43\,s & 0.51\,s & 0.89\,s & 1.56\,s & 4.43\,s\\
relative to Algorithm A & 1.1$\times$ & 1.3$\times$ & 2.3$\times$ & 4.0$\times$ & 11.2$\times$\\
\bottomrule
\end{tabular}
\end{adjustbox}
\end{table}

The surface alone is a weaker rule. Deep-searching the points with $\xi_1=\xi_2=0$ recovers $71\%$ of the genuine misses at $5.3\%$ of the
points, against $84\%$ at $5\%$ for the classifier, and adding the top $5\%$ of the other points by a classifier trained off the surface still does
not beat the classifier alone at the same budget ($88\%$ against $89\%$ at $10\%$ of the points; \cref{tab:shellsurface}): the classifier ranks the
dangerous points of the surface first, while the rule also spends budget on its safe half and on the cusp line. At $5<H\le6$ the surface holds only
$23\%$ of the genuine misses (\cref{sec:general}). The surface is a diagnosis rather than a ranker.

The locus of \cref{sec:double} recovers more than the surface, and at larger heights more than the classifier, at a higher cost (\cref{tab:aplus}). On its own,
deep-searching every point with $\xi_1^3=6\xi_2^2$ finds $726$ of the $836$ misses below height $5$, searching $17\%$ of the points ($14\%$
weighted by the census); with the repair it recovers $92\%$ of the misses, against $89\%$ for the classifier, and the gap widens at the larger
heights, where the classifier weakens. The classifier remains the more economical rule: it searches $5\%$ of the points at every height, the
exact rule about twice as many above height $5$. The exact rule needs no training, but it was found after the held-out samples had been evaluated
(\cref{sec:found}), and below height $7$ it selects, besides the surface, only two values of $(\xi_1,\xi_2)$ (\cref{sec:double}), so how well it
keeps selecting the hard instances at larger heights is untested. Searching where either rule selects recovers slightly more, at about the cost of
the exact rule.

The learning curve of the genuine-miss classifier decelerates but has not flattened: trained on $25$, $50$, $75$ and $100\%$ of the training folds it
reaches AUC $0.92$, $0.93$, $0.95$ and $0.95$.

\subsection{Generalisation to larger heights}
\label{sec:general}
At $5<H\le6$, we enumerated the census to $H\le6$ ($60\,661\,421$ points, $42\,804\,136$ fine and $78\,762$ with automorphisms,
\citealp[as in][Table~2]{m3}), and ran Algorithm~A, the reference search and the targeted repair on $10\,000$ points drawn uniformly at random, by
reservoir sampling, from the $29\,779\,198$ fine generic points with $5<H\le6$ and $\xi_0m_{12}\neq0$ (median $\halg=108$); $1.83\%$ of them lie on
the surface, against $1.91\%$ in the census, and its two halves are balanced ($97$ and $84$ points). The reference search finds a smaller model than Algorithm~A for $130$ of them ($1.3\%$),
$117$ genuinely ($1.2\%$), and its answer is smaller than the reference's best on one point. These rates are lower bounds that weaken with the height:
$33\%$ of the points have a prime above $13$ dividing the resultant to the second power, at which the search does not move, against $22\%$ below
$H=5$. The surface holds $27$ of the $117$ genuine misses ($23\%$), with a rate of $15\%$ on it and $0.92\%$ off it, as the two trends of
\cref{tab:shellsurface} predict. The classifiers trained on $H\le5$ only, applied without retraining, rank the genuine misses with AUC $0.94$
$[0.92,0.96]$ and AP $0.24$ at a base rate of $1.17\%$ (a $20$-fold enrichment), all misses with AUC $0.91$ $[0.88,0.94]$, and the misses off the
surface with AUC $0.90$ $[0.87,0.93]$. The top $5\%$ of the evaluation points by the classifier's score contains $72\%$ of the genuine misses and its
top $10\%$ contains $82\%$; the surface rule finds $87$ of the $117$ while searching $6.7\%$ of the points, and $89$ while searching $11.7\%$.
Algorithm~A$^+$ reaches the best model of the reference search on $82\%$ of the misses and $84\%$ of the genuine ones (\cref{tab:aplus}). The locus of multiple fixed points holds $88$ of the $117$ genuine misses, on
$9.4\%$ of the points (\cref{tab:multiplicity}).

For $6<H\le7$, the second held-out range, we enumerated the census to $H\le7$ ($214\,260\,288$ points, \citealp[as in][]{m3}), kept a random
$0.1\%$ of the points with $H>6$, and ran Algorithm~A, the reference search and the targeted repair on $10\,000$ random fine generic points with
$\xi_0m_{12}\neq0$ among them (median $\halg=144$). The method, including the classifier, its threshold and the move set of the repair, was fixed
before this shell was evaluated, and nothing was retrained. The reference search finds a smaller model for $188$ points ($1.9\%$), $166$ of them
genuine misses ($1.7\%$); $4.3\%$ of the searches reached the node budget, $34\%$ of the points have a prime above $13$ at which the search does not
move, and Algorithm~A's answer is smaller than the reference's best on three points. The surface holds $1.5\%$ of the points and $33$ of the $166$ genuine
misses ($20\%$), with rates of $22\%$ on it and $1.4\%$ off it. The classifiers rank the genuine misses with AUC $0.88$ $[0.85,0.91]$ (AP $0.17$ at a
base rate of $1.7\%$), all misses with AUC $0.88$ $[0.85,0.91]$ and the misses off the surface with AUC $0.80$ $[0.76,0.84]$. Algorithm~A$^+$ reaches
the best model of the reference search on $78\%$ of the misses and $83\%$ of the genuine ones (\cref{tab:aplus}). The locus of multiple fixed points holds $132$ of the $166$ genuine misses, on $8.4\%$ of the
points. As the height grows, the classifier weakens, and the repair and the exact rule, which need no training, carry more of the weight.

\subsection{Running time}
\label{sec:runtime}
Using per-point measurements of all its components on the same $1\,000$ random points of the main sample and of the uniform sample at $5<H\le6$, on
one machine with eight workers (\cref{tab:benchmark}), we reconstruct the running time of Algorithm~A$^+$ under the classifier's fixed selection rule:
it takes $1.8$ times the running time of Algorithm~A at $H\le5$ and $1.6$ times at $5<H\le6$, against $11$ and $7.9$ times for the depth-$5$ search on
every point. Of the extra time, the targeted repair accounts for $0.25$\,s per point, the features and the classifier for under a millisecond, and the
deep search on the flagged points for the rest. With the exact rule the cost depends on the share of points on the locus: $2.6$ times the running time of Algorithm~A at $H\le5$, where $18\%$
of the benchmark points lie on it, and $1.5$ times at $5<H\le6$, where $8.6\%$ do.

\begin{table}[ht]
\centering
\caption{Running time per point in seconds (mean, median, $90$th percentile) on $1\,000$ random points of each sample, with eight
workers. Algorithm~A includes step 5 as published; the other component rows start from the model after step 4; the rows for Algorithm~A$^+$
are reconstructed from the component times of each point.}
\label{tab:benchmark}
\small\setlength{\tabcolsep}{4pt}
\begin{adjustbox}{max width=\textwidth}
\begin{tabular}{lcccccc}
\toprule
 & \multicolumn{3}{c}{$H\le5$} & \multicolumn{3}{c}{$5<H\le6$}\\
\cmidrule(lr){2-4}\cmidrule(lr){5-7}
 & mean & median & p90 & mean & median & p90\\
\midrule
Algorithm~A & 0.448 & 0.282 & 0.863 & 0.615 & 0.401 & 1.344\\
\quad of which step 5 as published & 0.018 & 0.015 & 0.025 & 0.018 & 0.015 & 0.025\\
first-improvement step 5, for comparison & 0.017 & 0.015 & 0.022 & 0.017 & 0.014 & 0.022\\
targeted repair & 0.246 & 0.242 & 0.293 & 0.244 & 0.240 & 0.293\\
features and classifier & $<0.001$ & & & $<0.001$ & & \\
depth-$5$ reference search & 4.656 & 0.516 & 10.043 & 4.215 & 0.518 & 9.835\\
\midrule
Algorithm~A$^+$ & 0.818 & 0.575 & 1.194 & 0.955 & 0.696 & 1.645\\
Algorithm~A$^+$ with the exact rule & 1.166 & 0.703 & 1.287 & 0.949 & 0.748 & 1.677\\
Algorithm~A and the deep search everywhere & 5.104 & 0.854 & 11.098 & 4.830 & 1.044 & 10.495\\
\bottomrule
\end{tabular}
\end{adjustbox}
\end{table}

\section{Discussion and conclusion}
\label{sec:conclusion}

For one to two percent of the points we tested, a model smaller than the one found by exact algebra followed by a search of depth three lies
within reach of a deeper search. These hard instances are not noise. They concentrate on an exactly defined locus, the maps with a multiple fixed
point, where the fixed-point quartic is not stable: at small heights mostly on the maps with a triple fixed point, and there on the branch with real
attracting directions, and as the height grows increasingly on those with a double one. A certified structured search, Algorithm~A$^+$, finds most
of them, between $78\%$ and $92\%$ depending on the selection rule and the height range (\cref{tab:aplus}), at $1.5$ to $2.6$ times the running
time of Algorithm~A.

Two lessons go beyond these numbers. The first concerns the search. The improvements that Algorithm~A misses lie just beyond the radius of its
local search, and on the surface they share a recurring algebraic shape. The structure of its moves appears to matter more than its
radius: $8\,400$ structured composite moves recover more genuine misses than the reference search stopped at depth $4$, a larger search with more
moves, although the two are not a controlled comparison of work.

The second concerns learning. As in generate-and-verify methods \citep{funsearch,alphageometry}, exact verification is what makes the learned component safe here: since every candidate is checked by recomputing its
invariants, a classifier can decide where to spend computation with no risk to correctness, and its errors cost time or a missed improvement, never
a wrong answer. The same classifiers, interrogated, pointed to where the hard instances lie.

\subsection{Limitations}
The results have four kinds of limitation.
\begin{enumerate}
\item \emph{The reference.} A hard instance is defined relative to the reference search, a search of depth $5$ with primes up to $13$, which did
not complete on $4\%$ of the points and does not move at the primes above $13$ that divide the resultant to the second power, present at $22\%$ of
the points below height $5$ and at $33\%$ and $34\%$ in the held-out samples. The measured rates are therefore lower bounds for the rate at which
$\halg>\hmin$; only $461$ census points can be compared with exact values, and $\halg=\hmin$ on all of them.
\item \emph{The samples.} The main sample reaches $H\le5$ and the held-out samples $H\le7$, the largest height of the census \citep{m3}; exact
labels stop at $\hmin=5$; all data are over $\Q$; and the points with $\xi_0m_{12}=0$, $15\%$ of the fine generic points, are excluded because
Algorithm~A does not apply to them in its present form. Since the hard instances concentrate on a degenerate locus, these points, which form
another one, may well be harder than those we study.
\item \emph{The implementation.} $\halg$ depends on implementation choices in the heuristic steps: the two readings of step 5 compared in
\cref{sec:checks} miss on $836$ and $1\,646$ points. We follow the published description, but measured the dependence of $\halg$ on the starting
model only for the first-improvement reading. The boosted trees are not tuned.
\item \emph{The status of the explanations.} The locus of multiple fixed points is exact and has no fitted parameter, but it was tested after the
held-out samples had been evaluated, so its held-out figures, unlike the classifier's, are not a test fixed in advance, and its relation to the hard
instances is an observation rather than a theorem. That the reduction step is not the main cause rests on one experiment: adding a stable covariant
to step 4 removes only about a sixth of the hard instances on the locus (\cref{tab:step4}).
\end{enumerate}

\subsection{Open questions and future work}
This paper leaves several questions open. Our data show where the hard instances lie, but not why: we do not explain why smaller models are
harder to reach at multiple fixed points, why on the surface the branch with real attracting directions needs the elliptic correction $ST$, or why
the asymmetry between the branches fades with the height; and our methods give upper bounds for $\hmin$, not certificates of minimality. On the side
of learning, the classifier weakens as the height grows and its learning curve has not flattened, so how far its ranking carries beyond the heights
tested is open; the exact rule has been tested only on the two values of $(\xi_1,\xi_2)$ that occur below height $7$; and what the invariants
determine about the size of the model itself, rather than about where it is hard to find, is a separate question, which we leave to further work.

In the reduction step, adding the covariant point of the period-two form on the locus of multiple fixed points, as is done when the fixed-point
form is not stable \citep[\S6]{hutz-stoll}, removes about a sixth of the hard instances and creates very few (\cref{tab:step4}). Our data suggest
that the remaining hard instances call for a better search rather than a better reduction. For the search, a cost-aware policy could decide per point whether
to run the targeted repair, the deep search, or neither; a learned ordering of the composite moves, trained on the recorded paths together with the
moves that did not help, could act inside the search rather than before it; and a classifier on exact labels could predict the primes dividing the
conjugation from the algorithm's answer to a model of minimal height. Larger heights, beyond the census \citep{m3}, are the natural test of all
of these, and the same approach could be tried for rational maps of higher degree.

\section*{Data and code availability}
The programs, the notebook that recomputes every number, table and figure in this paper, and the result
set it writes are in the repository, together with Algorithm~A$^+$ as a single function and a check that it reproduces the stored evaluation \citep{minmodels}; the per-class labels, the census points and samples, and the outputs of Algorithm~A, of the reference search with its improving paths and of the targeted repair with its verified models are in its release \texttt{data-v1}. The conics of the descent are solved with \texttt{qfsolve} of PARI/GP \citep{pari}, through its Python interface, and the classifiers are those of scikit-learn \citep{sklearn}.

\appendix
\section{Algorithm A}
\label{app:algA}
Algorithm~A is the algorithm \citep[Alg.~7.1, arXiv version~1]{m3}, implemented as specified there. Moves are conjugations $\phi\mapsto\gamma^{-1}\circ\phi\circ\gamma$, with the result written as a primitive integer vector; $h$ is the naive height. The covariant points of step 4 are the roots in the upper half plane of the definite ones among the covariant quadratics $J$,
$(I,J)_2$, their Jacobian $u_3$ and their pairwise sums, and, when the fixed-point quartic $I$ has no repeated root, a point built from its roots
$\alpha_j$ \citep[Alg.~7.1(4)]{m3}: the minimiser in the upper half plane of
\[
\sum_jt_j\,\frac{|z-\alpha_j|^2}{\operatorname{Im}z},\qquad t_j=|I'(\alpha_j)|^{-1}.
\] 

When $I$ has a
repeated root, that is, on the locus of multiple fixed points of \cref{sec:double}, the point is not used and the current model is kept; our
implementation decides this exactly, by the discriminant of $I$. The points are computed with $60$ significant digits when a coefficient of the
model exceeds $10^9$, as prescribed.

\begin{algorithm}[H]
\caption{Algorithm~A \citep[Alg.~7.1]{m3}}\label{alg:A}
\begin{algorithmic}[1]
\Require a fine moduli point $P$ with trivial automorphism group and $\xi_0m_{12}\neq0$
\State $\phi_0\gets$ a map over $\Q$ with invariants $P$, from a rational point of the descent conic \citep[Prop.~3.7]{m3}
\Comment{step 1, exact}
\For{each prime $p$ with $p^6\mid\Res(\phi_0)$} \Comment{step 2, exact}
  \State conjugate by $\left(\begin{smallmatrix}p&a\\0&1\end{smallmatrix}\right)$ ($0\le a<p$) or $\left(\begin{smallmatrix}1&0\\0&p\end{smallmatrix}\right)$ while $|\Res|$ decreases
\EndFor
\State $\mathcal L\gets$ the models of the same resultant within distance $3$ in the trees at the primes $p$ with $p^2\mid\Res$ \Comment{step 3}
\For{each $\psi\in\mathcal L$} \Comment{step 4}
  \For{each covariant point $z$ of $\psi$}
    \State conjugate $\psi$ by the $M\in\SL_2(\Z)$ that moves $z$ into the standard fundamental domain
  \EndFor
  \State descend greedily under $G=\{z\mapsto z\pm1,\ -1/z,\ -z,\ z/(z\pm1)\}$, then try all words of length $2$ to $4$ in $G$
\EndFor
\State $\phi\gets$ the smallest model found so far
\Repeat \Comment{step 5}
  \State $\phi'\gets$ the smallest model within depth $3$ of $\phi$ under the eleven moves $z\mapsto z\pm1$, $-1/z$, $-z$,
  $\left(\begin{smallmatrix}p&a\\0&1\end{smallmatrix}\right)$ ($0\le a<p$) and $\left(\begin{smallmatrix}1&0\\0&p\end{smallmatrix}\right)$, $p=2,3$
  \If{$h(\phi')<h(\phi)$} $\phi\gets\phi'$ \EndIf
\Until{no model within depth $3$ is smaller}
\State \Return the smallest model found \Comment{step 6}
\end{algorithmic}
\end{algorithm}

Our implementation reads the heuristic steps as follows.
\begin{enumerate}
\item \emph{Step 5} keeps the smallest model of the whole depth-$3$ neighbourhood and repeats while the height decreases, as published. A
first-improvement reading, which moves at the first depth containing a smaller model, misses twice as often at the same cost (\cref{sec:checks}).
\item \emph{The covariant points of step 4} are used independently: for each point, the model of step 3 is conjugated by the matrix $M$ moving
that point into the fundamental domain and by $M^{-1}$, and the smallest of all these models is kept, rather than moving the points one after
another; the instruction to repeat until no covariant point moves \citep[Alg.~7.1(4)]{m3} is not implemented, the greedy descent under $G$ being
repeated while the height decreases instead.
\item \emph{The words of step 4} are applied to the model after the greedy descent and are not composed with the improvements they find: the search
is over all words of length at most $4$ from that model.
\item \emph{The point built from the roots of $I$} is computed in closed form as $z=x+iy$ with
\[
x=\frac{\sum_jt_j\operatorname{Re}\alpha_j}{\sum_jt_j},\qquad y=\Bigl(\frac{\sum_jt_j|\alpha_j-x|^2}{\sum_jt_j}\Bigr)^{1/2},
\]
which is the minimiser above for any roots $\alpha_j$, and is always among the candidates when $I$ has no repeated root.
\end{enumerate}

\bibliographystyle{elsarticle-harv}
\bibliography{references}

\begin{thebibliography}{29}
\expandafter\ifx\csname natexlab\endcsname\relax\def\natexlab#1{#1}\fi
\providecommand{\url}[1]{\texttt{#1}}
\providecommand{\href}[2]{#2}
\providecommand{\path}[1]{#1}
\providecommand{\DOIprefix}{doi:}
\providecommand{\ArXivprefix}{arXiv:}
\providecommand{\URLprefix}{URL: }
\providecommand{\Pubmedprefix}{pmid:}
\providecommand{\doi}[1]{\href{http://dx.doi.org/#1}{\path{#1}}}
\providecommand{\Pubmed}[1]{\href{pmid:#1}{\path{#1}}}
\providecommand{\bibinfo}[2]{#2}
\ifx\xfnm\relax \def\xfnm[#1]{\unskip,\space#1}\fi
\bibitem[{Badr et~al.(2027)Badr, Shaska and Shaska}]{jsc}
\bibinfo{author}{Badr, E.}, \bibinfo{author}{Shaska, E.},
  \bibinfo{author}{Shaska, T.}, \bibinfo{year}{2027}.
\newblock \bibinfo{title}{Rational functions on the projective line from a
  computational viewpoint}.
\newblock \bibinfo{journal}{Journal of Symbolic Computation}
  \bibinfo{volume}{140}, \bibinfo{pages}{102640}.
\newblock \DOIprefix\doi{10.1016/j.jsc.2026.102640}.
\bibitem[{Bengio et~al.(2021)Bengio, Lodi and Prouvost}]{bengio-lodi-2021}
\bibinfo{author}{Bengio, Y.}, \bibinfo{author}{Lodi, A.},
  \bibinfo{author}{Prouvost, A.}, \bibinfo{year}{2021}.
\newblock \bibinfo{title}{Machine learning for combinatorial optimization: a
  methodological tour d'horizon}.
\newblock \bibinfo{journal}{European Journal of Operational Research}
  \bibinfo{volume}{290}, \bibinfo{pages}{405--421}.
\bibitem[{Bruin and Molnar(2012)}]{bruin-molnar}
\bibinfo{author}{Bruin, N.}, \bibinfo{author}{Molnar, A.},
  \bibinfo{year}{2012}.
\newblock \bibinfo{title}{Minimal models for rational functions in a dynamical
  setting}.
\newblock \bibinfo{journal}{LMS Journal of Computation and Mathematics}
  \bibinfo{volume}{15}, \bibinfo{pages}{400--417}.
\newblock \DOIprefix\doi{10.1112/S1461157012001131}.
\bibitem[{Charton(2024)}]{charton}
\bibinfo{author}{Charton, F.}, \bibinfo{year}{2024}.
\newblock \bibinfo{title}{Learning the greatest common divisor: explaining
  transformer predictions}, in: \bibinfo{booktitle}{International Conference on
  Learning Representations (ICLR 2024)}.
\bibitem[{Cremona and Stoll(2003)}]{cremona-stoll}
\bibinfo{author}{Cremona, J.E.}, \bibinfo{author}{Stoll, M.},
  \bibinfo{year}{2003}.
\newblock \bibinfo{title}{On the reduction theory of binary forms}.
\newblock \bibinfo{journal}{Journal f{\"u}r die reine und angewandte
  Mathematik} \bibinfo{volume}{565}, \bibinfo{pages}{79--99}.
\bibitem[{Curri(2022)}]{curri}
\bibinfo{author}{Curri, E.}, \bibinfo{year}{2022}.
\newblock \bibinfo{title}{On the stability of binary forms and their weighted
  heights}.
\newblock \bibinfo{journal}{Albanian Journal of Mathematics}
  \bibinfo{volume}{16}, \bibinfo{pages}{3--20}.
\newblock \DOIprefix\doi{10.51286/albjm/1656410487}.
\bibitem[{Davies et~al.(2021)}]{davies-2021}
\bibinfo{author}{Davies, A.}, et~al., \bibinfo{year}{2021}.
\newblock \bibinfo{title}{Advancing mathematics by guiding human intuition with
  {AI}}.
\newblock \bibinfo{journal}{Nature} \bibinfo{volume}{600},
  \bibinfo{pages}{70--74}.
\bibitem[{Gasse et~al.(2019)Gasse, Ch{\'e}telat, Ferroni, Charlin and
  Lodi}]{gasse-2019}
\bibinfo{author}{Gasse, M.}, \bibinfo{author}{Ch{\'e}telat, D.},
  \bibinfo{author}{Ferroni, N.}, \bibinfo{author}{Charlin, L.},
  \bibinfo{author}{Lodi, A.}, \bibinfo{year}{2019}.
\newblock \bibinfo{title}{Exact combinatorial optimization with graph
  convolutional neural networks}, in: \bibinfo{booktitle}{Advances in Neural
  Information Processing Systems (NeurIPS 2019)}.
\bibitem[{Huang et~al.(2014)Huang, England, Wilson, Davenport, Paulson and
  Bridge}]{huang-2014}
\bibinfo{author}{Huang, Z.}, \bibinfo{author}{England, M.},
  \bibinfo{author}{Wilson, D.}, \bibinfo{author}{Davenport, J.H.},
  \bibinfo{author}{Paulson, L.C.}, \bibinfo{author}{Bridge, J.},
  \bibinfo{year}{2014}.
\newblock \bibinfo{title}{Applying machine learning to the problem of choosing
  a heuristic to select the variable ordering for cylindrical algebraic
  decomposition}, in: \bibinfo{booktitle}{Intelligent Computer Mathematics
  (CICM 2014)}, \bibinfo{publisher}{Springer}. pp. \bibinfo{pages}{92--107}.
\bibitem[{Hutz(2026)}]{hutz-ga}
\bibinfo{author}{Hutz, B.}, \bibinfo{year}{2026}.
\newblock \bibinfo{title}{A genetic algorithm for generating extreme examples
  in arithmetic dynamics}.
\newblock \bibinfo{note}{ArXiv:2601.11482}.
\bibitem[{Hutz and Stoll(2019)}]{hutz-stoll}
\bibinfo{author}{Hutz, B.}, \bibinfo{author}{Stoll, M.}, \bibinfo{year}{2019}.
\newblock \bibinfo{title}{Smallest representatives of
  {$\mathrm{SL}(2,\mathbb{Z})$}-orbits of binary forms and endomorphisms of
  {$\mathbb{P}^1$}}.
\newblock \bibinfo{journal}{Acta Arithmetica} \bibinfo{volume}{189},
  \bibinfo{pages}{283--308}.
\bibitem[{Janson(2011)}]{janson}
\bibinfo{author}{Janson, S.}, \bibinfo{year}{2011}.
\newblock \bibinfo{title}{Invariants of polynomials and binary forms}.
\newblock \href{http://arxiv.org/abs/1102.3568}{{\tt arXiv:1102.3568}}.
\bibitem[{Kotsireas and Shaska(2025)}]{kotsireas-shaska}
\bibinfo{author}{Kotsireas, I.}, \bibinfo{author}{Shaska, T.},
  \bibinfo{year}{2025}.
\newblock \bibinfo{title}{A neurosymbolic framework for geometric reduction of
  binary forms}.
\newblock \bibinfo{note}{ArXiv:2501.15404}.
\bibitem[{Kotthoff(2014)}]{kotthoff-2014}
\bibinfo{author}{Kotthoff, L.}, \bibinfo{year}{2014}.
\newblock \bibinfo{title}{Algorithm selection for combinatorial search
  problems: a survey}.
\newblock \bibinfo{journal}{AI Magazine} \bibinfo{volume}{35},
  \bibinfo{pages}{48--60}.
\bibitem[{Milnor(2006)}]{milnor-book}
\bibinfo{author}{Milnor, J.}, \bibinfo{year}{2006}.
\newblock \bibinfo{title}{Dynamics in One Complex Variable}. volume
  \bibinfo{volume}{160} of \textit{\bibinfo{series}{Annals of Mathematics
  Studies}}.
\newblock \bibinfo{edition}{Third} ed., \bibinfo{publisher}{Princeton
  University Press}.
\bibitem[{{minmodels}(2026)}]{minmodels}
\bibinfo{author}{{minmodels}}, \bibinfo{year}{2026}.
\newblock \bibinfo{title}{minmodels: code and data for ``{Where models of
  minimal height are hard to find}''}.
\newblock \bibinfo{howpublished}{GitHub repository,
  \url{https://github.com/LiraCurri/minmodels}}.
\newblock \bibinfo{note}{Release data-v1}.
\bibitem[{Pedregosa et~al.(2011)}]{sklearn}
\bibinfo{author}{Pedregosa, F.}, et~al., \bibinfo{year}{2011}.
\newblock \bibinfo{title}{Scikit-learn: machine learning in {P}ython}.
\newblock \bibinfo{journal}{Journal of Machine Learning Research}
  \bibinfo{volume}{12}, \bibinfo{pages}{2825--2830}.
\bibitem[{Peifer et~al.(2020)Peifer, Stillman and
  Halpern-Leistner}]{peifer-2020}
\bibinfo{author}{Peifer, D.}, \bibinfo{author}{Stillman, M.},
  \bibinfo{author}{Halpern-Leistner, D.}, \bibinfo{year}{2020}.
\newblock \bibinfo{title}{Learning selection strategies in {B}uchberger's
  algorithm}, in: \bibinfo{booktitle}{International Conference on Machine
  Learning (ICML 2020)}.
\bibitem[{Rice(1976)}]{rice-1976}
\bibinfo{author}{Rice, J.R.}, \bibinfo{year}{1976}.
\newblock \bibinfo{title}{The algorithm selection problem}.
\newblock \bibinfo{journal}{Advances in Computers} \bibinfo{volume}{15},
  \bibinfo{pages}{65--118}.
\bibitem[{Romera-Paredes et~al.(2024)}]{funsearch}
\bibinfo{author}{Romera-Paredes, B.}, et~al., \bibinfo{year}{2024}.
\newblock \bibinfo{title}{Mathematical discoveries from program search with
  large language models}.
\newblock \bibinfo{journal}{Nature} \bibinfo{volume}{625},
  \bibinfo{pages}{468--475}.
\bibitem[{Rumely(2015)}]{rumely}
\bibinfo{author}{Rumely, R.}, \bibinfo{year}{2015}.
\newblock \bibinfo{title}{The minimal resultant locus}.
\newblock \bibinfo{journal}{Acta Arithmetica} \bibinfo{volume}{169},
  \bibinfo{pages}{251--290}.
\bibitem[{Selsam et~al.(2019)Selsam, Lamm, B{\"u}nz, Liang, de~Moura and
  Dill}]{neurosat}
\bibinfo{author}{Selsam, D.}, \bibinfo{author}{Lamm, M.},
  \bibinfo{author}{B{\"u}nz, B.}, \bibinfo{author}{Liang, P.},
  \bibinfo{author}{de~Moura, L.}, \bibinfo{author}{Dill, D.L.},
  \bibinfo{year}{2019}.
\newblock \bibinfo{title}{Learning a {SAT} solver from single-bit supervision},
  in: \bibinfo{booktitle}{International Conference on Learning Representations
  (ICLR 2019)}.
\bibitem[{Shaska and Shaska(2025)}]{ss-jac}
\bibinfo{author}{Shaska, E.}, \bibinfo{author}{Shaska, T.},
  \bibinfo{year}{2025}.
\newblock \bibinfo{title}{Machine learning for moduli space of genus two curves
  and an application to isogeny-based cryptography}.
\newblock \bibinfo{journal}{Journal of Algebraic Combinatorics}
  \bibinfo{volume}{61}, \bibinfo{pages}{23}.
\newblock \DOIprefix\doi{10.1007/s10801-025-01393-8}.
\bibitem[{Shaska(2026)}]{m3}
\bibinfo{author}{Shaska, T.}, \bibinfo{year}{2026}.
\newblock \bibinfo{title}{Fine points of the moduli space of cubic rational
  maps}.
\newblock \href{http://arxiv.org/abs/2610.33990}{{\tt arXiv:2610.33990}}.
\bibitem[{Silverman(2007)}]{silverman-ads}
\bibinfo{author}{Silverman, J.H.}, \bibinfo{year}{2007}.
\newblock \bibinfo{title}{The Arithmetic of Dynamical Systems}. volume
  \bibinfo{volume}{241} of \textit{\bibinfo{series}{Graduate Texts in
  Mathematics}}.
\newblock \bibinfo{publisher}{Springer}, \bibinfo{address}{New York}.
\bibitem[{{The PARI Group}(2024)}]{pari}
\bibinfo{author}{{The PARI Group}}, \bibinfo{year}{2024}.
\newblock \bibinfo{title}{{PARI/GP}}.
\newblock \bibinfo{howpublished}{Univ. Bordeaux,
  \url{http://pari.math.u-bordeaux.fr/}}.
\newblock \bibinfo{note}{Version 2.15}.
\bibitem[{Trinh et~al.(2024)}]{alphageometry}
\bibinfo{author}{Trinh, T.H.}, et~al., \bibinfo{year}{2024}.
\newblock \bibinfo{title}{Solving olympiad geometry without human
  demonstrations}.
\newblock \bibinfo{journal}{Nature} \bibinfo{volume}{625},
  \bibinfo{pages}{476--482}.
\bibitem[{West(2014)}]{west}
\bibinfo{author}{West, L.W.}, \bibinfo{year}{2014}.
\newblock \bibinfo{title}{The moduli space of cubic rational maps}.
\newblock \href{http://arxiv.org/abs/1408.3247}{{\tt arXiv:1408.3247}}.
\bibitem[{Xu et~al.(2008)Xu, Hutter, Hoos and Leyton-Brown}]{satzilla}
\bibinfo{author}{Xu, L.}, \bibinfo{author}{Hutter, F.}, \bibinfo{author}{Hoos,
  H.H.}, \bibinfo{author}{Leyton-Brown, K.}, \bibinfo{year}{2008}.
\newblock \bibinfo{title}{{SATzilla}: portfolio-based algorithm selection for
  {SAT}}.
\newblock \bibinfo{journal}{Journal of Artificial Intelligence Research}
  \bibinfo{volume}{32}, \bibinfo{pages}{565--606}.

\end{thebibliography}

\end{document}